\documentclass[english]{revtex4-2}
\usepackage{textcomp}
\usepackage[utf8]{inputenc}
\usepackage{color}
\usepackage{babel}
\usepackage{array}
\usepackage{float}
\usepackage{booktabs}
\usepackage{mathtools}
\usepackage{multirow}
\usepackage{amsmath}
\usepackage{amssymb}
\usepackage{graphicx}
\usepackage{geometry}
\usepackage{setspace}
\usepackage[bookmarks=false,
 breaklinks=false,pdfborder={0 0 1},colorlinks=false]
 {hyperref}

\makeatletter

\newcommand{\lyxmathsym}[1]{\ifmmode\begingroup\def\b@ld{bold}
  \text{\ifx\math@version\b@ld\bfseries\fi#1}\endgroup\else#1\fi}

\providecommand{\tabularnewline}{\\}

\newcommand{\RNum}[1]{\uppercase\expandafter{\romannumeral #1\relax}}

\makeatother

\begin{document}
\title{Correlation-Driven Singlet--Triplet Inversion and Optical Spectra
of N-Substituted Phenalenyls: Insights from PPP-FCI and TDDFT}
\author{Medha Rakshit}
\author{Alok Shukla}
\email{shukla@iitb.ac.in}

\affiliation{Department of Physics, Indian Institute of Technology Bombay, Mumbai
400076, India}
\begin{abstract}
\noindent Several researchers have reported an atypical inversion of the lowest singlet and triplet excited states in nitrogen-substituted planar hydrocarbons. Contrary to Hund's rule of maximum multiplicity, poly-aza derivatives
of phenalenyl, such as cyclazine and heptazine, have long been known
to possess a negative singlet-triplet gap, $\Delta E_{ST}=E(S_{1})-E(T_{1})$,
where $S_{1}$ denotes the first singlet excited state, and $T_{1}$
the first triplet state. Using the well-known Pariser-Parr-Pople (PPP)
model, a semi-empirical, wave-function-based Hamiltonian for correlated $\pi$-electrons, we investigate
the singlet-triplet gap in ten N-substituted phenalenyl motifs. Two schemes of long-range Coulomb interactions are considered, and key parameters, including the on-site orbital energy, Hubbard U, and hopping integrals, are systematically tuned to benchmark the model for reliably estimating negative singlet-triplet gaps. Full configuration-interaction
(FCI) method is employed to incorporate electron-correlation effects.
Ground-state and $T_{1}$ absorption spectra are subsequently calculated using the optimized PPP parameters that best reproduce the available experimental data. For validation, first-principles TDDFT calculations using various exchange--correlation functionals are also performed. Conventional adiabatic TDDFT fails to reproduce the inverted ($S_{1}-T_{1}$) ordering, indicating that the inclusion of the correlation driven multiconfigurational effects are essential for describing it. We find that the singlet TDDFT ground-state absorption spectra reproduce several qualitative features of the PPP-FCI spectra, although the TDDFT excitation energies are systematically higher. Detailed analysis of the ground-state and $T_{1}$ absorption spectra reveals significant configuration mixing, providing a comprehensive characterization of the optical response of the
N-substituted phenalenyl derivatives studied.

\end{abstract}
\maketitle

\section{INTRODUCTION}

The ordering of the lowest singlet and triplet excited states is a
fundamental property governing the photophysics of molecular systems.
For a conventional closed-shell organic molecule, the lowest triplet
state ($T_{1}$) is expected to lie below the lowest singlet excited
state ($S_{1}$), consistent with the usual role of electron exchange
in determining the relative energies of different spin states. This
ordering underlies the conventional distinction between fluorescence,
associated primarily with the radiative ($S_{1}\rightarrow S_{0}$)
transition, and phosphorescence, which involves the spin-forbidden
($T_{1}\rightarrow S_{0}$) transition. When intersystem crossing
(ISC) is negligible, light emission mainly occurs as fluorescence,
from the lowest singlet excited state ($S_{1}$) to $S_{0}$ \citep{won2023inverted,de2019inverted}.
Research has focused on improving the internal quantum efficiency
(IQE) of organic light-emitting diodes (OLEDs) materials by harvesting
the remaining non-emissive triplet excitons. This effort has led to
the fabrication of organic compounds containing heavy-metal centers,
such as Ir \citep{lamansky2001molecularly}, Pt \citep{fleetham2017phosphorescent},
Os \citep{niu2005highly}, Ru \citep{buda2002thin}, etc., which enhance
the spin--orbit coupling and allow radiative decay from the triplet
state through phosphorescence. In molecular materials, the energy
difference between these states, $\Delta E_{ST}=E(S_{1})-E(T_{1})$,
therefore provides a particularly sensitive measure of the electronic
interactions governing low-lying excited states. Another emerging
class of materials called thermally activated delayed fluorescence
(TADF) emitters, with a narrow singlet-triplet gap (STG) $\Delta E_{ST}=E(S_{1})-E(T_{1})$
between $S_{1}$ and $T_{1}$ that allows up-conversion of nonradiative
triplet excitons, to radiative singlet excitons, through a reverse
intersystem crossing (RISC) \citep{li2022down,garner2023double}.
The comparatively long lifetime (in the range of $10^{-6}s$ to $10^{-3}s$)
of TADF emitters with positive $\Delta E_{ST}$ degrade their efficiency
\citep{won2023inverted,li2022down} due to the triplet-triplet annihilation
and triplet-polaron annihilation, which counteract the RISC. In this
context, the materials with a narrow negative $\Delta E_{ST}$, i.e.,
in which $S_{1}$ state is positioned lower in energy than the $T_{1}$
state, are more efficient as these materials undergo a down-conversion
process from $T_{1}$ to $S_{1}$ \citep{li2022down}. The microscopic
origin of singlet--triplet gap inversion has been associated with
various factors, including reduced exchange interactions associated
with spatially separated frontier orbitals, molecular resonance patterns,
electron correlation and dynamic spin polarization, while vibronic
and pseudo-Jahn--Teller interactions can further influence the relative
energetics of the low-lying states \citep{drwal2023role,pollice2024rational,dinkelbach2021large,majumdar2024influence}.
Such systems have consequently been termed inverted singlet-{}-triplet
(INVEST) emitters and have attracted attention as molecular systems
with unconventional excited-state dynamics and potentially useful
optical properties. Several classes of molecules exhibiting inverted
singlet--triplet energy ($\Delta E_{ST}<0$) have been identified
over the past several years, including azaphenalenes \citep{ehrmaier2019singlet,loos2023heptazine,pollice2021organic},
engineered non-alternant hydrocarbons such as pentalene and azulene
analogues \citep{majumdar2025unlocking,terence2023symmetry,garner2024enhanced,pollice2024rational},
isopyrene \citep{garner2023double}, calicene, bicalicene \citep{terence2023symmetry,blaskovits2024singlet},
boron--nitrogen embedded polyaromatic architectures \citep{bedogni2024singlet,pios2021triangular},
etc.. These molecular platforms have substantially expanded the structural
landscape for achieving negative STGs, providing promising candidates
for the development of next-generation INVEST materials. Importantly,
however, the sign and magnitude of ($\Delta E_{\mathrm{ST}}$) alone
do not determine the photophysical performance of a molecule. The
radiative accessibility of $S_{1}$, spin--orbit coupling, nonradiative
relaxation, and the electronic and vibronic character of the participating
states can all influence the observable optical response. The problem
is therefore more fundamental than identifying molecular structures
with small or negative singlet--triplet gaps; it concerns how molecular
topology and electron-correlation cooperate to produce an unconventional
ordering of correlated many-electron states. Thus, understanding the
microscopic origin of singlet--triplet inversion is important not
only for molecular design but also as a fundamental problem in excited-state
electronic-structure theory. 

A particularly well-defined molecular family in which singlet--triplet
inversion can be investigated is provided by azaphenalenes, which
are obtained through systematic substitution of carbon sites in the
phenalenyl framework by nitrogen atoms. Phenalenyl is an alternant,
odd $\pi$-electrons system with a doublet ground state; however,
replacement of its central carbon atom by nitrogen produces cycl{[}3.3.3{]}azine
(cyclazine), an even $\pi$-electrons system that exhibits the characteristic
inverted singlet--triplet ordering. Leupin \emph{et al.} were the
first to experimentally predict that cyclazine can deviate from Hund’s
rule, and result in a negative STG \citep{leupin1980low}. Further
substitution of the peripheral carbon sites by nitrogen generates
a series of polyazaphenalene derivatives that retain the underlying
13 atom and 14 $\pi$-electron framework as cyclazine \citep{rossman1985synthesis,hosmane1982synthesis,lindqvist19781}.
This chemically controlled substitution provides a systematic means
of modifying the electronic structure while maintaining a closely
related molecular backbone, making azaphenalenes particularly useful
model systems for examining the singlet--triplet inversion and optical
accessibility of their $S_{1}$state. In another experimental study,
Leupin \emph{et al.} revealed a negative STG in triazacycl{[}3.3.3{]}azine
and heptazine \citep{Leupin1986}. Aizawa \emph{et al.} experimentally
observed a negative STG of -11$\pm$2 meV in a heptazine analogue
molecule \citep{aizawa2022delayed}. Li \emph{et al.} reported an
STG of -0.22 eV and -0.19 eV for 2,5,8-tris (4-fluoro-3-methylphenyl)heptazine
in toluene and acetonitrile, respectively \citep{li2022down}. An
experimental investigation on 2,5,8‑Tris(phenylthiolato)heptazine
by Blasco \emph{et al.} also reported a negative STG \citep{blasco2024experimental}.
In dialkylamine-substituted pentazaphenalene, Kusakabe et al. experimentally
measured a negative STG of -41 meV and -32 meV in the sample film
and sample solution, respectively \citep{kusakabe2024inverted}. In
a recent experimental study using high-resolution cryogenic anion
photoelectron spectroscopy, Wilson \emph{et al.} measured an STG of
-47 meV in pentazaphenalene \citep{wilson2024spectroscopic}. The
combination of reduced spatial overlap of orbitals \citep{won2023inverted,garner2023double},
partial charge-transfer, multi-configurational character of the first
singlet excited state \citep{ehrmaier2019singlet}, modified aromatic
electron distribution, and enhanced electron-correlation effects shifts
the balance in favor of the lowering the energy of $S_{1}$, resulting
in a negative STG \citep{dreuw2023inverted}. 

Theoretical investigations of singlet--triplet inversion in azaphenalenes
have employed a broad range of electronic-structure methods, including
configuration-interaction, perturbative, coupled-cluster, multireference,
and time-dependent density-functional theory (TDDFT) approaches \citep{won2023inverted,de2019inverted,drwal2023role,ehrmaier2019singlet,loos2023heptazine,pollice2021organic,blasco2024experimental,ricci2021singlet,dreuw2023inverted}.
Although correlated wave-function methods such as ADC(2), CC2, EOM-CCSD,
and CASPT2 have been reported to reproduce negative $\Delta E_{ST}$,
linear-response TDDFT has generally struggled to describe the inversion,
with the predicted $\Delta E_{ST}$ depending sensitively on the treatment
of electron correlation. This difficulty is particularly relevant
because the low-lying excited states may possess substantial double-excitation
or multiconfigurational character, which is not adequately captured
within conventional linear-response TDDFT. Given the widespread use
of TDDFT for molecular excitation energies and optical spectra, a
systematic comparison with an explicitly correlated treatment in a
controlled molecular model is therefore important for assessing the
reliability of TDDFT and elucidating the electronic origin of singlet--triplet
inversion.

To gain deeper insight into the physics driving ST inversion in azaphenalenes,
and to report a detailed study of their ground-state and $T_{1}$
absorption spectra, we have chosen a total of ten azaphenalene structures
including AP1, i.e., cycl{[}3.3.3{]}azine ($C_{12}H_{9}N$); AP2,
i.e., 2-monoazacycl{[}3.3.3{]}azine ($C_{11}H_{8}N_{2}$); two isomers
of ($C_{10}H_{7}N_{3}$), AP31, i.e., 6,7-diazacycl{[}3.3.3{]}azine,
and AP32, i.e., 5,8-diazacycl{[}3.3.3{]}azine; two isomers of ($C_{9}H_{6}N_{4}$),
AP41, i.e., 1,3,6-triazacycl{[}3.3.3{]}azine, and AP42, i.e., 1,3,4-triazacycl{[}3.3.3{]}azine;
AP5, i.e., 1,3,4,-tetraazacycl{[}3.3.3{]}azine ($C_{8}H_{5}N_{5}$);
two isomers of ($C_{7}H_{4}N_{6}$), AP61, i.e., 1,3,4,6,7-pentaazacycl{[}3.3.3{]}azine,
and AP62, i.e., 1,3,4,6,8-pentaazacycl{[}3.3.3{]}azine; and AP7, i.e.,
1,3,4,6,7,9-hexaazacycl{[}3.3.3{]}azine, tri-s-triazine, or heptazine
($C_{6}H_{3}N_{7}$). This work employs the Pariser--Parr--Pople (PPP) Hamiltonian, a
wave-function-based semi-empirical effective $\pi$-electron model
introduced in the 1950s \citep{pariser-parr1953semi,pople1953electron}.
To examine the dependence of singlet--triplet inversion on the treatment
of long-range electron--electron interactions, both the Ohno and
Mataga-{}-Nishimoto forms of the Coulomb interaction were considered
\citep{ohno1964,mataga1957}. For each interaction form, the PPP parameters,
including the on-site orbital energies, Hubbard $U$, and hopping
matrix elements, are systematically varied, and a total of six parameter
sets are examined to benchmark the model against available experimental
results and to assess its ability to reproduce negative STG of AP1.
The excited-state properties of the molecules were further investigated
using TDDFT with several exchange--correlation functionals and full
configuration interaction (FCI) within the PPP framework, enabling
a direct assessment of their respective abilities to describe singlet--triplet
inversion. In particular, negative $\Delta E_{\mathrm{ST}}$ values
are obtained for the relevant azaphenalene systems with four sets
of PPP-parameterization, whereas the TDDFT calculations considered
here yield positive STGs. This parameter sensitivity is particularly
relevant for heteroatom-containing systems, where differences in the
local orbital energies and on-site Coulomb interactions at carbon
and nitrogen sites influence the electronic structure in addition
to the intersite Coulomb interaction. Beyond energetic inversion,
we also examine the optical accessibility of $S_{1}$, since a negative
$\Delta E_{\mathrm{ST}}$ alone does not guarantee a strong optical
response. Several azaphenalene derivatives exhibiting negative $\Delta E_{\mathrm{ST}}$
possess a nearly dark $S_{1}$ state, whereas others show appreciable
$S_{0}\rightarrow S_{1}$ oscillator strengths, demonstrating that
the energetic ordering, many-electron character, and optical transition
strength of the low-lying excited states are distinct but complementary
characteristics of an INVEST system. Finally, using the PPP-FCI method
with the OHNO-\RNum{3} parameterization, we calculated the ground-state
and $T_{1}$ absorption spectra of the molecules exhibiting negative
STGs, providing a comprehensive characterization of their optical
properties and the effects of nitrogen substitution across the possible
N-substituted phenalenyl motifs within the PPP framework. The novelty
of the present work lies in providing a unified, systematic description
of the electronic structure and optical properties of ten N-substituted
phenalenyl-derived azaphenalenes within a correlated-electron framework.
Negative STGs are newly predicted for AP41 and AP61, while the ($T_{1}\rightarrow T_{n}$)
absorption spectra are reported for the first time for the negative-STG
members of the series. In contrast to the limited singlet-state spectroscopic
information available for this family, where complete absorption spectra
have been reported only for AP1, AP5, and AP7, and previous theoretical
studies of AP2, AP31, AP32, and AP42 have primarily considered the
($S_{0}\rightarrow S_{1}$) transition, the present study systematically
resolves the higher singlet excited states across the molecular series
exhibiting negative STG. The combination of PPP-FCI calculations,
multiple electron--electron interaction parameterizations, and complementary
TDDFT calculations further allows us to distinguish energetic singlet--triplet
inversion from the correlated character and optical accessibility
of the corresponding excited states. 

\section{COMPUTATIONAL METHODOLOGY}

There are many established computational approaches to explore the
electronic and optical properties of $\pi$-conjugated nanostructures
which we broadly \textcolor{black}{classify} into two classes: (a)
first-principles approaches such as density-functional theory (DFT)
which is a mean-field approach, or those based on the many-electron
wave function such as the configuration interaction (CI), coupled-cluster
(CC) approach, etc., and (b) model Hamiltonian based methods such
as Dirac-equation-based massless Fermion approach, effective $\pi$-electron
models such as the H\"uckel model, which is nothing but an independent-electron
tight-binding (TB) approach, and its electron-correlated extensions
such as the Hubbard, extended Hubbard, or the Pariser-Parr-Pople models
\citep{gundra2013shukla}. In this work, we employed the Pariser-Parr-Pople
(PPP) model \citep{pariser-parr1953semi,pople1953electron}, which
goes beyond the Hubbard model, and includes long-range electron-electron
interactions. We used the CI methodology to obtain the electron-correlated
wave functions of the ground and excited states, which were subsequently
employed to compute the optical response of the systems under consideration.
We also performed first-principles calculations employing the time-dependent
DFT (TDDFT) approach to compute the optical response of those molecules,
and compare the results with those obtained using the PPP model based
approach. Next, we briefly describe the PPP and TDDFT approaches.

\begin{figure}[!t]
\centering \includegraphics[width=6in,totalheight=4in,keepaspectratio,height=5.5in]{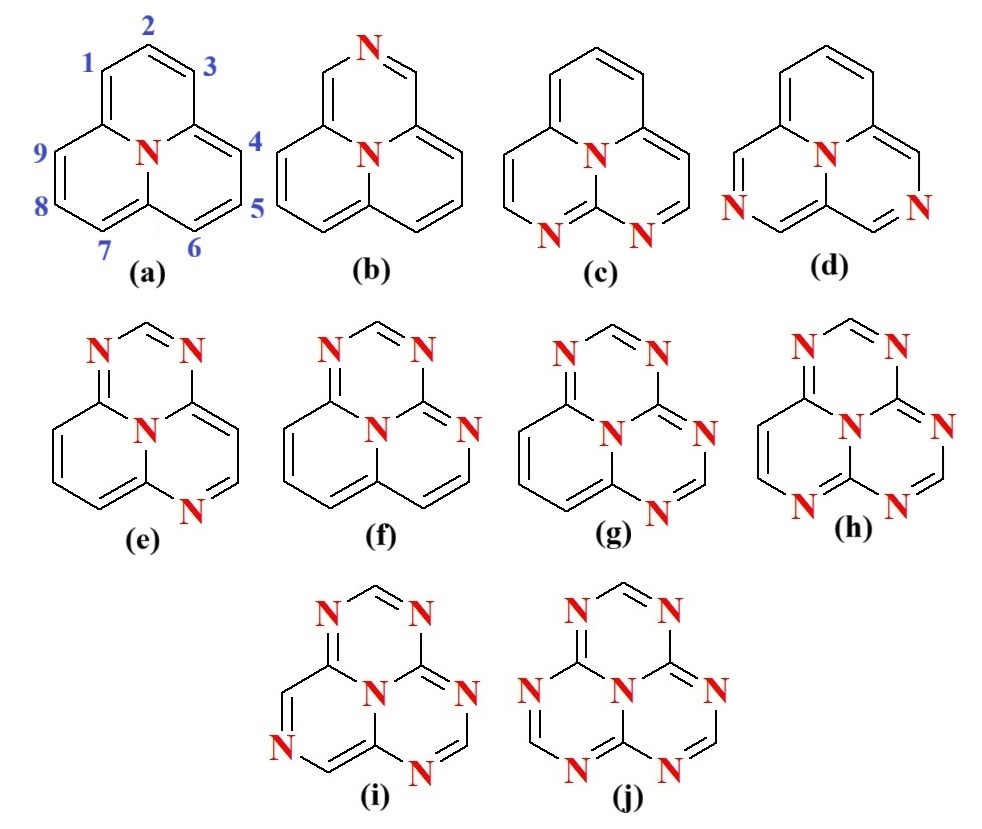}
\caption{Representation of APs studied in the present work, (a) AP1, i.e.,
cycl{[}3.3.3{]}azine, (b) AP2, i.e., 2-monoazacycl{[}3.3.3{]}azine,
(c) AP31, i.e., 6,7-diazacycl{[}3.3.3{]}azine, (d) AP32, i.e., 5,8-diazacycl{[}3.3.3{]}azine,
(e) AP41, i.e., 1,3,6-triazacycl{[}3.3.3{]}azine, (f) AP42, i.e.,
1,3,4-triazacycl{[}3.3.3{]}azine, (g) AP5, i.e., 1,3,4,-tetraazacycl{[}3.3.3{]}azine,
(h) AP61, i.e., 1,3,4,6,7-pentaazacycl{[}3.3.3{]}azine, (i) AP62,
i.e., 1,3,4,6,8-pentaazacycl{[}3.3.3{]}azine, and (j) AP7, i.e., 1,3,4,6,7,9-hexaazacycl{[}3.3.3{]}azine,
Tri-s-triazine, or Heptazine.}
\label{Fig.1.} 
\end{figure}

\subsection{ PPP Model-Based Approach}

The PPP model assumes that the electronic and optical properties of
$\pi$-conjugated systems (here heterocyclic and aromatic) can be
accurately described by the behavior of their $\pi$-electrons, with
the core and $\sigma$-electrons treated as inert. These lower-energy
electrons lie far from the Fermi level and remain unaffected by external
perturbations, with their effects incorporated in the model parameters.
Assuming that the molecule under consideration lies in the $xy$-plane,
each atom which is a part of the conjugation network, contributes
a specific number of $\pi$-electrons populating a localized $p_{z}$
orbital, forming an orthonormal basis, under the zero-differential
overlap (ZDO) approximation\citep{parr1952method}. The PPP Hamiltonian
is thus formulated in second quantization form as follows

\begin{equation}
\widehat{H_{PPP}}=\sum_{i}\varepsilon_{i}\hat{n}_{i}-\sum_{ij,i<j}\sum_{\sigma}t_{ij}(\hat{c}_{i\sigma}^{\dag}\hat{c}_{j\sigma}+\hat{c}_{j\sigma}^{\dag}\hat{c}_{i\sigma})+\sum_{i}U_{i}\hat{n}_{i\uparrow}\hat{n}_{i\downarrow}+\sum_{ij,i<j}V_{ij}(Z_{i}-\hat{n}_{i})(Z_{j}-\hat{n}_{j}),
\end{equation}

where, $\hat{c}_{i\sigma}(\hat{c}_{i\sigma}^{\dag})$ annihilates
(creates) an electron of spin $\sigma$ on the $i^{th}$ atomic site,
$Z_{i}$ is the total number of $\pi$ electrons contributed by the
$i$-th atom, and $\hat{n}_{i}=\sum_{\sigma}\hat{c}_{i\sigma}^{\dag}\hat{c}_{i\sigma}$
sum up the total number of $\pi$-electrons on the $i^{th}$ atomic
site. The four parameters, $\varepsilon_{i}$, $t_{ij}$, $U_{i}$,
and $V_{ij}$ used in the PPP Hamiltonian are referred to as the on-site
orbital energy, hopping amplitude between $i^{th}$ and $j^{th}$
atomic sites, on-site Coulombic repulsion between two electrons on
the $i^{th}$ atom, and the long-range Coulombic repulsion, respectively.
Here, we have considered the hopping matrix elements beyond the nearest-neighbor
to be zero. In order to compute $t_{ij}$ in eV units, we used the
expression\citep{das-ramasesha}

\begin{equation}
t_{ij}=t_{0}+3.20(r_{ij}-r_{0}),
\end{equation}
where $r_{ij}$ is the distance between the $i$-th and $j$-th sites
in $\text{\AA}$ units, and $r_{0}$ denotes a reference bond length
for which the hopping integral takes the value $t_{0}$.

As far as the long-range Coulomb interaction $V_{ij}$ is concerned,
we considered two forms, the one proposed by Ohno \citep{ohno1964},
and another by Mataga and Nishimoto \citep{mataga1957}. The form
of $V_{ij}$ for homoatomic $\pi$-conjugated system, as proposed
by Ohno is given by,

\begin{equation}
V_{ij}=\frac{U_{i}}{\left[1+\left(\frac{r_{ij}}{r_{0}}\right)^{2}\right]^{1/2}}.\label{eq:ohno}
\end{equation}

For heteroatomic systems such as the ones considered in this work,
this formula is generalized as\citep{das2018low,thomas2013linear} 

\begin{equation}
V_{ij}=\frac{U_{ij}}{\left[1+\left(\frac{r_{ij}U_{ij}}{14.39964}\right)^{2}\right]^{1/2}},\label{eq:ohno-hetero}
\end{equation}

where $U_{ij}=(U_{i}+U_{j})/2$. Moreover, to incorporate the screening
effects in our calculations, we adopt a slightly modified formulation
proposed by Chandross and Mazumdar \citep{chandross1997coulomb} for
carbon-based conjugated materials, with the dielectric constant $\kappa_{ij}=2,\:\mbox{ for }i\neq j$,
while $\kappa_{ii}=1$

\begin{equation}
V_{ij}=\frac{U_{ij}}{\kappa_{ij}\left[1+\left(\frac{r_{ij}U_{ij}}{14.39964}\right)^{2}\right]^{1/2}}.
\end{equation}

We also generalized the formulation of Mataga and Nishimoto\citep{mataga1957}
to compute $V_{ij}$, with the possibility of including the screening
effects

\begin{equation}
V_{ij}=\frac{U_{ij}}{\kappa_{ij}\left[1+\left(r_{ij}U_{ij}/14.39964\right)\right]}
\end{equation}

\subsubsection{Choice of the PPP Parameters}

In this work, we have considered ten N-doped phenalenyl derivatives,
ranging from a single dopant N atom up to seven N atoms, and their
basic structures are shown in Figure ~\ref{Fig.1.}(a)---(j). In
a given structure, each vertex without a label denotes a carbon atom,
while nitrogen atoms are indicated by label N. Each carbon atom on
the edge is also assumed attached to a hydrogen atom. All our previous
works employing the PPP model were on planar hydrocarbons\citep{gundra2013shukla},
including the recent ones\citep{pttp-arifa,t-graphene-arifa,das2024dft}.
Therefore, without any heteroatom, the choice of parameters was rather
simple: (a) $\epsilon_{i}=0.0$ eV, $U_{i}=11.13$ eV, and $\kappa_{ij}=1.0$
called the standard parameters, and (b) (a) $\epsilon_{i}=0.0$ eV,
$U_{i}=8.0$ eV, $\kappa_{ii}=1.0$, and $\kappa_{ij}=2.0$, for $i\neq j$,
called screened parameters.\citep{chandross1997coulomb} Because of
the planar nature of the systems, it was assumed that each carbon
atom contributes only one $\pi$ electron, so that $Z_{i}=1$, and
only the Ohno parameterization scheme (Eq. \ref{eq:ohno}) was used. 

However, in the present work, because of the presence of N atoms,
we have to provide PPP parameters for both C and N atoms. Furthermore,
we have also performed calculations using the Mataga-Nishimoto parameterization
scheme. As for planar hydrocarbons, each C atom contributes one $\pi$
electron to the system ($Z_{i}=1$), while the nitrogen atoms bonded
to three atoms (pyrrolic nitrogens) contribute two $\pi$ electrons
($Z_{i}=2$), and those bonded to two atoms (aza nitrogens) contribute
one $\pi$ electron ($Z_{i}=1$). 

In summary, we have tested six sets of PPP parameters (see Table \ref{tab:ppp-parameter})
on AP1 based on the form of $V_{ij}$ (Ohno or Mataga-Nishimoto) used
in the Hamiltonian, and the values $U_{i}$, and $\epsilon_{i}$.
We name these parameters as OHNO-\RNum{1}, OHNO-\RNum{2}, OHNO-\RNum{3},
MN-\RNum{1}, MN-\RNum{2}, and MN-\RNum{3}. For all other molecules
(AP2 - AP7), we have used only those parameter sets that correctly
predicted the negative singlet-triplet gap of AP1, i.e. OHNO-\RNum{2},
\RNum{3}; and MN-\RNum{2}, \RNum{3}. A detailed description
of all the parameter sets is provided in Table ~\ref{tab:ppp-parameter}.
However, the\textcolor{black}{{} STG }of the remaining molecules calculated
using the MN-\RNum{1} and MN-\RNum{2} parameterizations are significantly
larger than the experimentally measured values. Moreover, the OHNO-\RNum{2}
parameterization fails to predict the \textcolor{black}{STG} reliably
for molecules containing peripheral nitrogen atoms (such as AP2--AP7).
Therefore, we restrict our subsequent spectral calculations to the
OHNO-\RNum{3} parameterization. Previous experimental and theoretical
studies on many azaphenalenes have revealed that these molecules violate
Hund’s multiplicity rule in the sense that their first singlet excited
state ($S_{1}$) is lower in energy compared to the first triplet
state ($T_{1}$). However, to our knowledge, no computational analysis
of the singlet and triplet optical spectra on the complete class of
azaphenalenes is currently available. In this work, we fill that gap,
and compute both the singlet and triplet excitation spectra of the
molecules using the OHNO-\RNum{3} parameterization, as this choice
yields results in good agreement with the available experimental spectra,
as discussed comprehensively in section \ref{subsec:PPP-CI-Results}. 

\begin{table}[th]
\centering
\caption{Parameter sets considered for the PPP model Hamiltonian  in this
work. The superscripts \emph{aza} and \emph{py }, respectively, imply
the aza and pyrrolic types of bonds involving the nitrogen atoms.}\label{tab:ppp-parameter}
\smallskip{}
\begin{tabular}{ccccccccc}
\toprule 
\multirow{3}{*}{\textcolor{black}{$V_{ij}$}} & \multirow{3}{*}{\textcolor{black}{Name}} & \multicolumn{6}{c}{Parameters} & \multirow{3}{*}{Ref}\tabularnewline
\cmidrule(l){3-8}
 &  & \multicolumn{3}{c}{\textcolor{black}{$U_{i}$}} & \multicolumn{3}{c}{\textcolor{black}{$\epsilon_{i}$}} & \tabularnewline
\cmidrule{3-8}
 &  & \textcolor{black}{$U_{C}$} & \textcolor{black}{$U_{N}^{py}$} & \textcolor{black}{$U_{N}^{aza}$} & \textcolor{black}{$\epsilon_{C}$} & \textcolor{black}{$\epsilon_{N}^{py}$ } & \textcolor{black}{$\epsilon_{N}^{aza}$ } & \tabularnewline
\midrule 
\multirow{3}{*}{\textcolor{black}{Ohno}} & \textcolor{black}{OHNO-\RNum{1}} & 11.13 & 14.80 & - & -11.16 & -19.72 & - & \citep{HinzeJaffe1962}\tabularnewline
 & \textcolor{black}{OHNO-\RNum{2}} & 11.26 & 15.00 & 15.00 & 0 & -12.00 & -2.00 & \citep{thomas2013linear}\tabularnewline
 & \textcolor{black}{OHNO-\RNum{3}} & 11.26 & 15.00 & 15.50 & 0 & -13.00 & -5.00 & \citep{bedogni2023shining}\tabularnewline
\multirow{3}{*}{\textcolor{black}{Mataga-Nishimoto}} & \textcolor{black}{MN-\RNum{1}} & 10.66 & 12.98 & - & 0 & -2.89 & - & \citep{mataga1957}\tabularnewline
 & \textcolor{black}{MN-\RNum{2}} & 11.26 & 15.00 & 15.00 & 0 & -12.00 & -2.00 & \citep{thomas2013linear}\tabularnewline
 & \textcolor{black}{MN-\RNum{3}} & 11.26 & 15.00 & 15.50 & 0 & -13.00 & -5.00 & \citep{bedogni2023shining}\tabularnewline
\bottomrule
\end{tabular}
\end{table}

\subsubsection{PPP-model Computational Workflow}

Our calculations are initiated through a restricted Hartree--Fock
(RHF) analysis tailored for the singlet ground states of the closed-shell
systems, given the fact that all molecules considered have fourteen,
i.e., an even number, of $\pi$ electrons each. For this purpose,
a Fortran 90 computer program developed in our group to perform PPP-model
calculations is utilized.\citep{ppp-model-shukla} The molecular orbitals
(MOs) thus obtained are utilized to transform the PPP Hamiltonian
from its original atomic orbital (AO) basis into the MO basis, which
sets the stage for further electron-correlation studies via the configuration
interaction (CI) method. Following this, we perform spin- and symmetry-adapted
full-CI (FCI) calculations using the MELD software package to determine
ground-state and optically accessible excited-state energies and their
corresponding many-electron wavefunctions \citep{MELD-code}. Transition
dipole moments are then calculated to quantify the optical transitions
from the ground state to the excited states, ultimately enabling the
calculation of the optical absorption spectrum, using the formula,

\begin{equation}
\sigma(\omega)=4\pi\alpha\sum_{n}\frac{\omega_{ng}|\langle n|\hat{e}.\mathbf{r}|g\rangle|^{2}\gamma^{2}}{\left(\omega_{ng}-\omega\right)^{2}+\gamma^{2}}.\label{eq:sigma}
\end{equation}

Above $|g\rangle$ represents the correlated ground-state CI wave
function, while $|n\rangle$ denotes the excited-state wave function
to which the optical transition occurs, with the energy difference
between the states $\hbar\omega_{ng}=E_{n}-E_{g}$. The term $\langle n|\hat{e}.\mathbf{r}|g\rangle$
denotes the transition dipole matrix element between the two states,
for the absorption of a photon of polarization direction $\hat{e}$,
and the energy $\hbar\omega_{ng}$. Finally, $\alpha$ is the fine-structure
constant, and $\gamma$ is a line width of the excited states assumed
to be uniform for all of them. Note that the expression of Eq. \ref{eq:sigma}
can also be used to compute the transitions from an excited states
by replacing $g$ by that particular excited state in the expression
above.

The oscillator strength $f$ of a specific transition is computed
as,

\begin{equation}
f=\frac{2m_{e}}{3\hbar^{2}}(E)\sum_{j=x,y,z}|\langle e|O_{j}|R\rangle|^{2},
\end{equation}

where $|R\rangle$, $|e\rangle$ denote the wave functions of the
two states, $E$ is the energy difference between them, and $O_{j}$
denotes the $j$-th Cartesian component of the dipole operator. 

\subsection{First-Principles DFT Approach}

All our first-principles calculations were performed within the general
framework of density functional theory (DFT), as implemented in Gaussian
16 \citep{gaussian16}. First, for each molecule considered in this
work, the geometry optimization was performed, followed by a vibrational
frequency analysis, to check its dynamical stability. For both sets
of calculations, we employed the restricted B3LYP hybrid exchange-correlation
functional \citep{LYP1988,Becke1988,Becke1993}, in conjunction with
the 6-31++G(d,p) double-zeta basis set \citep{Hehre1972}. \textcolor{black}{During
the self-consistent field (SCF) procedure, a convergence tolerance
of 1.0\texttimes 10\textsuperscript{-}\textsuperscript{6} Hartrees
was applied. For the geometry optimization, convergence was considered
achieved when the following criteria were met: a maximum atomic force
of 0.00045 Hartree/Bohr, an RMS force of 0.0003 Hartree/Bohr, a maximum
displacement of 0.0018 Bohr, and an RMS displacement of 0.0012 Bohr.}
The absence of imaginary vibrational frequencies confirms that all
optimized structures correspond to minima on the calculated potential-energy
surfaces. To obtain the optical absorption spectra of the azaphenalenes,
we employed the TDDFT approach, also implemented within Gaussian 16.
The TDDFT calculations of APs used a range of exchange-correlation
functionals, namely, Perdew--Burke--Ernzerhof (PBE) \citep{PBE},
B3LYP (which merges Becke’s three-parameter exchange with the Lee--Yang--Parr
correlation), CAM-B3LYP which is a long-range corrected variant of
B3LYP, applying the Coulomb-attenuating method\citep{CAMB3LYP}, and
the Heyd--Scuseria--Ernzerhof (HSE06) functional \citep{HSE06}.
The same basis set, 6-31++G(d,p), was used for the TDDFT calculations,
which is also employed in the ground-state optimizations.

\section{RESULTS AND DISCUSSION}

In this section we present and analyze the results of our calculations,
and it is organized as follows. First, we briefly discuss the optimized
geometry of each of the considered molecules, followed by their single-electron
energy levels at the DFT and PPP-HF levels of theories. Subsequently,
we present and compare the singlet-triplet ($S_{1}-T_{1}$) gaps of
the systems computed using the DFT and PPP-FCI approaches. Finally,
we present the optical absorption spectra of these molecules both
from their singlet ground state ($S_{0}$), and the first triplet
excited state, $T_{1}$. The ground-state absorption spectra are computed
using both the TDDFT and PPP-FCI approaches, while the $T_{1}$ absorption
spectra are computed only at the PPP-FCI level of theory. 

\subsection{Structural Optimization}

\textcolor{black}{For all the molecules, the geometries of the doubly-occupied
singlet ($S_{0}$) and the first triplet ($T_{1}$) configurations
were optimized using the first-principles DFT-based approach, and
in all cases, the singlet structure had energy lower than that of
the corresponding triplet one, and was identified as the $S_{0}$
ground state. Furthermore, the point-group symmetries of the ground
states of molecules AP1, AP2, AP31, AP32, AP41, AP42, AP5, AP61, AP62,
and AP7 were found to be $D_{3h}$, $C_{2v}$, $C_{2v}$, $C_{2v}$,
$C_{1}$, $C_{1}$, $C_{2v}$, $C_{1}$, $C_{2v}$, and $D_{3h}$,
respectively, and all of them have planar structures. The average
ground-state C-C and C-N bond lengths for all the ten molecules is
presented in Table S1 of Supplementary Material file.} We
note that both carbon-carbon and carbon-nitrogen (pyrrolic) bond lengths
are close to 1.40 \AA, while carbon-nitrogen (aza) bond lengths are
significantly smaller $\approx$ 1.33 \AA. We also computed the ground-state
vibrational frequencies of the considered molecules, and found all
of them to be real, implying that the obtained ground-state structures
are dynamically robust.

\subsection{One-electron Energy Levels}

\begin{figure}[H]
\centering \includegraphics[width=6.5in,totalheight=7in,height=7.2in]{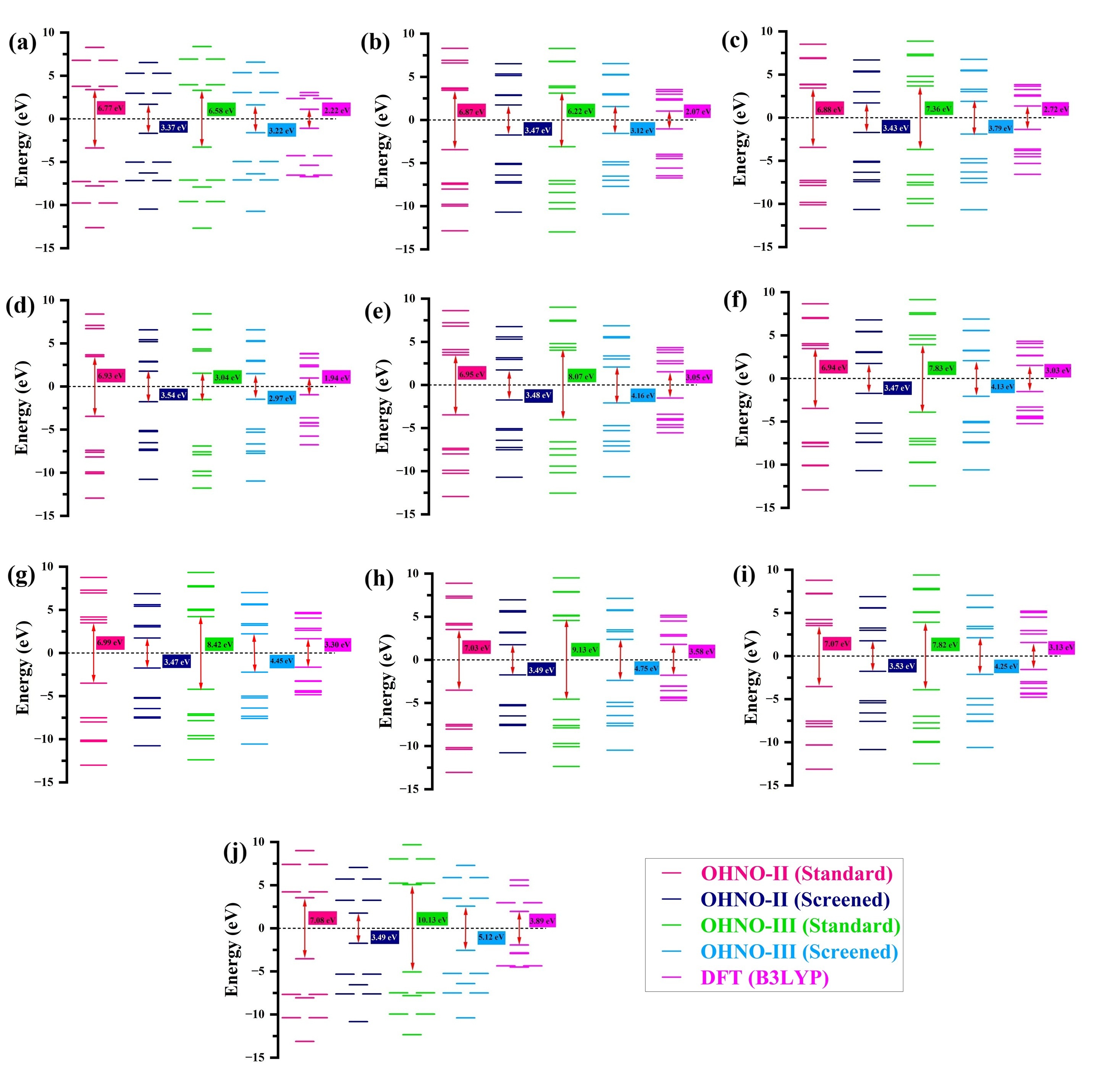}
\caption{Schematic diagram of the energy levels corresponding to HOMO-6 to
LUMO+5 of (a) AP1, (b) AP2, (c) AP31, (d) AP32, (e) AP41, (f) AP42,
(g) AP5, (h) AP61, (i) AP62, and (j) AP7 computed using the PPP-HF
and DFT calculations. The Fermi level of all the systems is set to
0 eV to compare the calculated HOMO-LUMO energy gaps using different
approaches easily. }
\label{Fig: energy-levels} 
\end{figure}

We begin by briefly describing the electronic structure of the azaphenalenes
(as given in Figure ~\ref{Fig.1.}) at the one-electron level, analyzed
using both DFT and PPP-Hartree-Fock (PPP-HF) methods, based on their
optimized geometries. Figure ~\ref{Fig: energy-levels} illustrates
the electronic energy levels of orbitals near the Fermi level (from
$H-6$ to $L+5$), computed using DFT and PPP-HF approaches. Here,
OHNO-\RNum{2} and OHNO-\RNum{3} parameterizations are employed
under standard and screened conditions to compute the energy levels
using the PPP-HF method. All the methods predict doubly degenerate
$H-1$, $H-3$, $L+1$, and $L+2$ energy levels in AP1 and AP7, consistent
with their $D_{3h}$ point-group symmetry. As shown in Figure ~\ref{Fig: energy-levels},
the energy level arrangements from both computational approaches are
reasonably consistent across all molecules. However, the HOMO-LUMO
gap, $E_{HL}$, is the largest for PPP-HF method computed using the
standard parameters, and the smallest when computed using DFT (see
also Table S2).\textcolor{blue}{{} }The smaller DFT Kohn--Sham
HOMO--LUMO gaps compared with the PPP-HF orbital gaps reflect the
well-known limitations of the Kohn--Sham orbital energy differences
as quantitative estimates of fundamental excitation or electronic
gaps. Using the PPP-HF approach under standard and screened parameter\textcolor{black}{{}
scheme}s, the lowest value of $E_{HL}$ is observed in AP1, while
using DFT (B3LYP), the lowest gap is observed in AP2. The highest
$E_{HL}$ value is observed in AP7, using the standard PPP parameters,
and also DFT.

\subsection{$S_{1}-T_{1}$ Gaps}

In this section we present the results of our calculations on the
\textcolor{black}{STGs} of azaphenalenes computed using the TDDFT
and PPP-FCI approaches. In TDDFT calculations B3LYP, HSE06, CAM-B3LYP,
and PBE exchange-correlation functionals were used. 

\subsubsection{TDDFT Results}

The calculated \textcolor{black}{STGs} of the azaphenalene molecules
employing the TDDFT method with different exchange-correlation functionals
(B3LYP, HSE06, CAM-B3LYP, PBE), coupled with the 6-31++G(d,p) basis
set are listed in \textcolor{black}{Table S3}. We note that
in all the cases, the TDDFT calculations predict $S_{1}$ to be higher
than $T_{1}$, a result in disagreement with the PPP-FCI calculations.
Our results on AP1 and AP7 predicting $S_{1}>T_{1}$, are consistent
with the TDDFT calculations of other authors on those molecules \citep{de2019inverted,ehrmaier2019singlet},
thereby giving us confidence about the correct implementation of this
methodology in Gaussian16. This outcome underscores a fundamental
limitation of standard linear-response TDDFT when paired with conventional
adiabatic exchange-correlation functionals. Because conventional adiabatic
linear-response TDDFT describes excited-state response primarily through
single-particle transition space and therefore cannot adequately represent
states with substantial double-excitation and multiconfigurational
character driving the inverted $S_{1}-T_{1}$ ordering. This behavior
clearly is an inherent limitation of conventional adiabatic functionals,
rather than a systemic failure of the broader TDDFT framework itself.
Indeed, specialized modern variants, such as spin-flip TDDFT (SF-TDDFT)
\citep{shao2003spin}or double-hybrid functionals \citep{grimme2007double}
have been shown in recent literature to occasionally resolve these
complex, correlation-driven states. Within the scope of standard adiabatic
approaches, however, explicitly correlated wave-function models remain
essential.

\subsubsection{PPP-FCI Results}

\label{subsec:PPP-CI-Results} 

In this section, first we briefly discuss the nature of FCI wave functions
of $S_{0}$, $S_{1}$, and $T_{1}$ states of the molecules studied
in this work. Next, to benchmark our results against those of other
authors and experiments, we compare our FCI results with those in
the literature for AP1, AP5, and AP7, followed by a brief discussion
of the remainder of the molecules.

\subsubsection*{The many-particle nature of $S_{0}$, $S_{1}$, and $T_{1}$ States}

We have presented the numerically significant configurations contributing
to the many-particle wave functions of the ground state ($S_{0}$),
and the excited states $S_{1}$ and $T_{1}$ in Table S35 of the SI.
We note that for all the molecules except AP32 the $S_{0}$ wave function
is dominated by the HF configuration, with the coefficient close to
0.9 or more. (b) AP32 is a peculiar case for which with the dominant
configuration of $S_{0}$ is the double excitation {\footnotesize$|H\rightarrow L;H\rightarrow L\rangle,$}
with coefficient 0.86, along with small contributions from other double
excitations. The fact that HF mean-field configuration does not contribute
to the FCI ground state of AP32, implies that this system is strongly
correlated, and that HF state is not able to represent its ground
state accurately. Further investigation revealed that the total energy
of the $|H\rightarrow L;H\rightarrow L\rangle$ is lower than that
of the $|HF\rangle$ configuration, even though the single-particle
HF energy of LUMO is higher than that of HOMO. In other words, for
this case, Aufbau principle based on the single-particle HF energies
is not a good descriptor of the ground state. However, there are no
conceptual problems due to this peculiar result, because our calculations
for all the molecules are based on the FCI method, which is exact
within the chosen model.

As far as $S_{1}$ wave functions are concerned, for all the molecules
they are dominated by the singly-excited configuration $|H\rightarrow L\rangle$,
with small contributions from other configurations. When it comes
to $T_{1}$ wave functions, we find that for all the molecules except
AP7, $|H\rightarrow L\rangle$ makes the dominant contributions to
the corresponding FCI wave functions. However, for AP7, the situation
is slightly different in that we find a doubly degenerate $T_{1}$
state with the dominant configurations $|H\rightarrow(L+1)_{1}\rangle$
and $|H\rightarrow(L+1)_{2}\rangle$, where the virtual orbitals $|(L+1)_{1}\rangle$
and $|(L+1)_{2}\rangle$ are degenerate. In Table S35, we have given
the $T_{1}$ FCI wave function corresponding to the dominant configuration
$|H\rightarrow(L+1)_{1}\rangle$, while the one corresponding to $|H\rightarrow(L+1)_{2}\rangle$
can be deduced from symmetry arguments. 

\subsubsection*{Comparison of Results for AP1 }

We initiated the calculations by computing the \textcolor{black}{STGs}
of AP1 using six sets of Coulomb parameters in the PPP model, listed
in Table ~\ref{tab:GAP_PPP}, and Table S4 under FCI level of CI
theory. Our results show that the FCI calculations performed using
the OHNO-\RNum{2}, \RNum{3}, MN-\RNum{2}, and MN-\RNum{3}
parameters predict a negative\textcolor{black}{{} STG} value for AP1.
The estimated STG values for AP1 using the FCI level theory with OHNO-\RNum{2}
(OHNO-\RNum{3})and MN-\RNum{2} (MN-\RNum{3}) parametrization
are -0.147 (-0.193) eV and -0.358 (-0.408) eV, respectively. Pollice
et al. estimated the STG of AP1 to be -0.160, -0.161, and -0.144 eV
using double-hybrid time-dependent density function approximation
(TD-DFA) calculations, such as ADC(2)/cc-pVDZ, ADC(2)/cc-pVDZ/IEFPCM(S0),
and ADC(2)/aug-cc-pVDZ, respectively. Using spin-flip TD-DFA, i.e.,
SA-SF-PBE50/def2-SVP, they estimated the gap to be -0.109 eV \citep{pollice2021organic}.
Loos \emph{et al.} reported STG of AP1 to be -0.137, -0.130, and -0.131
eV employing ADC(2)/aug-cc-pVTZ, CC2/aug-cc-pVTZ, and CC3/aug-cc-pVDZ
method, respectively \citep{loos2023heptazine}. EOM-CCSD/cc-pVDZ
calculation level produces an STG of -0.093 eV in AP1 \citep{garner2024enhanced}.
The previously reported STG values and the vertical energy positions
of the $S_{1}$ and $T_{1}$ states of AP1 obtained using different
theoretical methods are in good agreement with our FCI/OHNO-\RNum{2}
and FCI/OHNO-\RNum{3} results. The existing experimental data for
AP1 report $S_{0}-S_{1}$ and $S_{0}-S_{2}$ excitation gaps of 0.97
eV and 2.71 eV, respectively, whereas the present work obtains corresponding
values of 0.92 eV and 2.65 eV using FCI/OHNO-\RNum{3} parametrization.
\citep{leupin1980low}. 

\subsubsection*{Comparison of Results for AP5:}

Because our AP1 calculations using the MN-\RNum{2}, and MN-\RNum{3}
parameterizations did not show good agreement with the results reported
in other works, for AP5, we restrict the comparison of our PPP-FCI
calculations performed only using the OHNO-\RNum{2} and OHNO-\RNum{3}
parameters. Using these parameters, the STGs predicted by our PPP-FCI
calculations are -0.163 and -0.017 eV using the OHNO-\RNum{2}, and
OHNO-\RNum{3} parameters, respectively (Table \ref{tab:GAP_PPP}).
Additionally, our PPP-FCI $S_{0}-S_{1}$ gaps are 1.0888 (2.2485)
eV with\textcolor{red}{{} }OHNO-\RNum{2} (OHNO-\RNum{3}) parameters\textcolor{red}{.}
Wilson \emph{et al.} reported an experimental STG of -0.047 eV and
an $S_{0}-S_{1}$ excitation gap of 1.96 eV \citep{wilson2024spectroscopic},
indicating that our OHNO-\RNum{3} parameter results are in better
agreement with the experiments, as compared to those computed using
the OHNO-\RNum{2} parameters. In the same study by Wilson et al.,
EOM-CCSDt/cc-pVTZ//B3LYP predicted an $S_{0}-S_{1}$ gap of 1.942
eV and an STG of -0.022 eV, while using EOM-CCSDt/cc-pVTZ the obtained
values of the same are 2.223 eV and -0.092 eV, respectively \citep{wilson2024spectroscopic}.
However, employing EOM-CCSD/cc-pVTZ method the obtained $S_{0}-S_{1}$
gap and STG are 2.343 eV and -0.030 eV, respectively. Loos et al.
estimated $S_{0}-S_{1}$ gap of 2.374 eV and an STG of -0.020 eV \citep{loos2023heptazine}.
In another study, EOM-CCSD/cc-pVDZ level of theory, estimated the
$S_{0}-S_{1}$ gap of 2.251 eV and an STG of -0.078 eV, which also
aligns well with our results \citep{pollice2021organic}. 

\subsubsection*{Comparison of Results for AP7: }

Our calculated STG values of AP7 are -0.185 eV and -0.217 eV, using
the OHNO-\RNum{2}, and OHNO-\RNum{3} parameterizations, respectively,
which are in good agreement with the STG values obtained in other
works, such as -0.180 eV (EOM-CCSD/cc-pVDZ), -0.188 eV ($\omega$B2PLYP$^{\prime}$/def2-SV(P)
(adiabatic)), -0.181 eV (SA-SF-PBE50/def2-SVP), -0.167 eV (FNO-EOM-CCSD/aug-cc-pVDZ),
-0.198 eV ($\omega$B2PLYP$^{\prime}$/def2-SV(P) (vertical)), -0.214
eV (FNO-EOM-CCSD/cc-pVDZ), -0.182 eV (CCSD/cc-pVDZ), -0.139 eV ( SC-NEVPT2/def2-TZVP),
-0.130 eV (RASPT2/def2-TZVP), -0.166 eV (EOM-CCSD/cc-pVDZ), etc. \citep{pollice2021organic,loos2023heptazine,garner2024enhanced}.
AP7 shows an experimental $S_{0}-S_{1}$ gap of 2.60 eV, which is
reasonably well reproduced by the FCI/OHNO-\RNum{3} calculation
that gives 2.90 eV. In contrast, the FCI/OHNO-\RNum{2} parametrization
predicts a much lower value of 1.03 eV, showing poor agreement with
the experiment. Loos et al. theoretically estimated $S_{0}-S_{1}$
gap of 2.717 eV, and STG of -0.219 eV employing CC3/aug-cc-pVTZ+{[}CCSDT/6-31+G(d)\textminus CC3/6-31+G(d){]}
for $S_{1}$, and CC3/aug-cc-pVDZ+{[}CCSD/aug-cc-pVTZ\textminus CCSD/aug-cc-pVDZ{]}
for $T_{1}$\citep{loos2023heptazine}, in excellent agreement with
our values. Therefore, we conclude that for the PPP-FCI calculations,
OHNO-\RNum{3} parameters are most suited for all the molecules considered
in this work.

\subsubsection*{Results for the rest of the molecules}

To the best of our knowledge, there is no experimental evidence reported
for the STG gap in the remaining molecules (AP2, AP31, AP32, AP41,
AP42, AP61, and AP62). However, theoretical data for AP2, AP31, AP32,
and AP42, computed using different levels of theory, is available.
Using CC3/aug-cc-pVTZ+{[}CCSDT/6-31+G(d)\textminus CC3/6-31+G(d){]}
scheme for $S_{1}$ and CC3/aug-cc-pVDZ+{[}CCSD/aug-cc-pVTZ\textminus CCSD/aug-cc-pVDZ{]}
scheme for $T_{1}$, Loos et al. reported STG of -0.071, -0.101, and
-0.042 eV for AP2, AP31, and AP32, respectively, whereas, our work
estimates those as -0.120, -0.091, and -0.099 eV, respectively \citep{loos2023heptazine}.
Our analysis reveals an $S_{0}-S_{1}$ gap of 1.919 eV and an STG
of -0.039 eV for AP42 using FCI/OHNO-\RNum{3} parameterization,
close to the predicted $S_{0}-S_{1}$ gap of 2.012 eV and an STG of
-0.029 eV under EOM-CCSD/cc-pVDZ level of theory \citep{pollice2021organic}.
We have observed negative STG for AP41 and AP61 using both FCI/OHNO-\RNum{2}
and \RNum{3} schemes. For AP62, FCI/OHNO-\RNum{3} yields a positive
STG, whereas FCI/OHNO-\RNum{2} gives a negative value. In the absence
of experimental or theoretical STG data for AP41, AP61, and AP62 we
rely on the FCI/OHNO-\RNum{3} result, as this parametrization best
reproduces the available experimental data for AP1, AP5, and AP7.
Calculated STG for AP1-AP7 using FCI/OHNO-\RNum{2} and \RNum{3}
parametrization scheme are listed in Table \ref{tab:GAP_PPP}, and
the STG result using other parameter schemes are listed in Table S4. We hope that, in future, calculations and experimental measurements
will be performed on AP41, AP61, and AP62, against which our PPP-FCI
results can be tested. 

\begin{table}
\centering
\caption{The vertical excitation energies of $S_{1}$ and $T_{1}$ states with
respect to the ground state ($S_{0})$ for various APs, along with
their singlet-triplet ($S_{1}-T_{1}$) gaps. Additionally, the oscillator
strengths ($f$) corresponding to the $S_{0}-S_{1}$ and $S_{0}-S_{2}$
transitions are also presented. All the calculations were performed
using the FCI method coupled with the OHNO-\RNum{2} and OHNO-\RNum{3}
parameterizations, and the reported energy gaps are in the eV units.
}\label{tab:GAP_PPP}
\smallskip{}

{\scriptsize{}%
\begin{tabular}{ccccccc}
\toprule 
\multirow{2}{*}{{\scriptsize\textbf{System}}} & \multirow{2}{*}{{\scriptsize\textbf{Method}}} & \multicolumn{5}{c}{{\scriptsize\textbf{FCI}}}\tabularnewline
\cmidrule(l){3-7}
 &  & {\footnotesize\textbf{$E(S_{0})-E(S_{1})$}} & {\footnotesize\textbf{$E(S_{0})-E(T_{1})$}} & {\footnotesize\textbf{$E(S_{1})-E(T_{1})$}} & {\footnotesize\textbf{$f(S_{0}-S_{1})$}} & {\footnotesize\textbf{$f(S_{0}-S_{2})$}}\tabularnewline
\midrule
\multirow{2}{*}{{\scriptsize AP1}} & {\scriptsize OHNO-\RNum{2}} & {\scriptsize 1.1354689} & {\scriptsize 1.2822778} & {\scriptsize -0.147} & {\scriptsize 0.0000} & {\scriptsize 0.1389}\tabularnewline
 & {\scriptsize OHNO-\RNum{3}} & {\scriptsize 0.9234578} & {\scriptsize 1.1163319} & {\scriptsize -0.193} & {\scriptsize 0.0000} & {\scriptsize 0.1241}\tabularnewline
\midrule
\multirow{2}{*}{{\scriptsize AP2}} & {\scriptsize OHNO-\RNum{2}} & {\scriptsize 1.2201126} & {\scriptsize 1.3744035} & {\scriptsize -0.154} & {\scriptsize 0.0000} & {\scriptsize 0.1353}\tabularnewline
 & {\scriptsize OHNO-\RNum{3}} & {\scriptsize 0.7691364} & {\scriptsize 0.8893851} & {\scriptsize -0.120} & {\scriptsize 0.0007} & {\scriptsize 0.0975}\tabularnewline
\midrule 
\multirow{2}{*}{{\scriptsize AP31}} & {\scriptsize OHNO-\RNum{2}} & {\scriptsize 1.1236708} & {\scriptsize 1.2832391} & {\scriptsize -0.160} & {\scriptsize 0.0003} & {\scriptsize 0.1079}\tabularnewline
 & {\scriptsize OHNO-\RNum{3}} & {\scriptsize 1.5523735} & {\scriptsize 1.6436559} & {\scriptsize -0.091} & {\scriptsize 0.0096} & {\scriptsize 0.1429}\tabularnewline
\midrule 
\multirow{2}{*}{{\scriptsize AP32}} & {\scriptsize OHNO-\RNum{2}} & {\scriptsize 1.2424414} & {\scriptsize 1.4079175} & {\scriptsize -0.165} & {\scriptsize 0.0000} & {\scriptsize 0.1402}\tabularnewline
 & {\scriptsize OHNO-\RNum{3}} & {\scriptsize 0.5546077} & {\scriptsize 0.6537141} & {\scriptsize -0.099} & {\scriptsize 0.0005} & {\scriptsize 0.0787}\tabularnewline
\midrule 
\multirow{2}{*}{{\scriptsize AP41}} & {\scriptsize OHNO-\RNum{2}} & {\scriptsize 1.1264549} & {\scriptsize 1.2819651} & {\scriptsize -0.156} & {\scriptsize 0.0001} & {\scriptsize 0.1278}\tabularnewline
 & {\scriptsize OHNO-\RNum{3}} & {\scriptsize 1.9106828} & {\scriptsize 2.0177436} & {\scriptsize -0.107} & {\scriptsize 0.0009} & {\scriptsize 0.1706}\tabularnewline
\midrule 
\multirow{2}{*}{{\scriptsize AP42}} & {\scriptsize OHNO-\RNum{2}} & {\scriptsize 1.1104058} & {\scriptsize 1.2670012} & {\scriptsize -0.157} & {\scriptsize 0.0001} & {\scriptsize 0.1091}\tabularnewline
 & {\scriptsize OHNO-\RNum{3}} & {\scriptsize 1.9188739} & {\scriptsize 1.9575395} & {\scriptsize -0.039} & {\scriptsize 0.0135} & {\scriptsize 0.1648}\tabularnewline
\midrule 
\multirow{2}{*}{{\scriptsize AP5}} & {\scriptsize OHNO-\RNum{2}} & {\scriptsize 1.0888292} & {\scriptsize 1.2517513} & {\scriptsize -0.163} & {\scriptsize 0.0000} & {\scriptsize 0.1099}\tabularnewline
 & {\scriptsize OHNO-\RNum{3}} & {\scriptsize 2.2485326} & {\scriptsize 2.2656327} & {\scriptsize -0.017} & {\scriptsize 0.0089} & {\scriptsize 0.1867}\tabularnewline
\midrule 
\multirow{2}{*}{{\scriptsize AP61}} & {\scriptsize OHNO-\RNum{2}} & {\scriptsize 1.0550385} & {\scriptsize 1.2306323} & {\scriptsize -0.176} & {\scriptsize 0.0001} & {\scriptsize 0.0941}\tabularnewline
 & {\scriptsize OHNO-\RNum{3}} & {\scriptsize 2.5414073} & {\scriptsize 2.5842600} & {\scriptsize -0.043} & {\scriptsize 0.0076} & {\scriptsize 0.1964}\tabularnewline
\midrule 
\multirow{2}{*}{{\scriptsize AP62}} & {\scriptsize OHNO-\RNum{2}} & {\scriptsize 1.1156357} & {\scriptsize 1.3036523} & {\scriptsize -0.188} & {\scriptsize 0.0000} & {\scriptsize 0.1249}\tabularnewline
 & {\scriptsize OHNO-\RNum{3}} & {\scriptsize 2.0361030} & {\scriptsize 1.8392619} & {\scriptsize +0.197} & {\scriptsize -} & {\scriptsize -}\tabularnewline
\midrule 
\multirow{2}{*}{{\scriptsize AP7}} & {\scriptsize OHNO-\RNum{2}} & {\scriptsize 1.0254222} & {\scriptsize 1.2108961} & {\scriptsize -0.185} & {\scriptsize 0.0000} & {\scriptsize 0.0757}\tabularnewline
 & {\scriptsize OHNO-\RNum{3}} & {\scriptsize 2.9090841} & {\scriptsize 3.1262623} & {\scriptsize -0.217} & {\scriptsize 0.0000} & {\scriptsize 0.2037}\tabularnewline
\bottomrule
\end{tabular}}{\scriptsize\par}
\end{table}

\clearpage{}

\subsection{$S_{0}-S_{1}$ Oscillator Strengths}

Before describing in detail the optical absorption spectra of the
\textcolor{black}{azaphenalenes} considered in this work, we discuss
their oscillator strengths ( $f$ ) for the lowest singlet transition
$S_{0}-S_{1}$ (see Table \ref{tab:GAP_PPP}), to assess their potential
for direct fluorescence-based OLED applications. We find that it is
zero up to four decimal places for AP1, AP7; and negligible for AP2,
AP32, AP41. This is significant because, it implies that in these
molecules, the lowest singlet excitation is practically optically
forbidden, i.e., a dark state. As a result, their fluorescence is
very weak, and despite the singlet-triplet inversion, these compounds
are less promising as direct fluorescence-based OLED emitters. Conversely,
we find reasonable oscillator strengths for AP31, AP42, AP5, and AP61,
in agreement with EOM-CCSD/cc-pVDZ results \textcolor{blue}{\citep{pollice2021organic}},
which also predict significant\textbf{ $f(S_{0}-S_{1})$} values for
AP31, AP42 and AP5. Of all the molecules studied, AP42 exhibits the
maximum $S_{0}-S_{1}$ oscillator strength $f(S_{0}-S_{1})=0.0135$,
which compares favorably with the EOM-CCSD/cc-pVDZ result of $f=0.0050$
reported by Pollice et al.\textcolor{black}{{} \citep{pollice2021organic}. }

\subsection{Ground-State Absorption }

Next, we report and discuss the ground-state absorption spectra of
azaphenalenes, computed using the first-principles TDDFT and the PPP-FCI
methods. The ground-state spin multiplicities of all the APs considered
were singlet, therefore, their ground-state absorption spectra were
computed in the singlet manifolds. Since AP62 does not exhibit a negative
STG with the selected OHNO- \RNum{3} parameters, its optical absorption
spectrum is not included in the subsequent analysis of the negative-STG
systems.

\subsubsection{TDDFT Spectra}

We observe a common trend in the TDDFT spectra of all molecules (Figure
S2). The hybrid functionals B3LYP and HSE06 predict nearly identical
first significant absorption peak position. In contrast, PBE predicts
the first significant peaks at lower energies with reduced intensities,
whereas CAM-B3LYP shifts these peaks to higher energies with generally
enhanced intensities. Moreover, unlike the other functionals, PBE
gives a substantially weaker and less resolved description of the
higher-energy spectral region compared with the hybrid functionals.
This limitation arises from the flawed asymptotic behavior of its
exchange-correlation (XC) potential and its poor description of diffuse
and charge-transfer excitations, making it unreliable for high-energy
transitions, although it remains reasonably accurate for localized
low-energy excitations due to partial error cancellation \citep{li2015improving,dreuw2004failure}.
On the other hand, due to its range-separated formulation combined
with long-range Hartree-Fock exchange contributions, CAM-B3LYP predicts
peaks shifting to higher energies relative to the other functionals,
and sometimes overestimates the optical gap \citep{colin2023theoretical,shao2019benchmarking,peach2008excitation}. 

In the SI, we provide a detailed peak analysis of the B3LYP/TDDFT
spectra for all the molecules. B3LYP is widely used for conjugated
$\pi$-electron systems and generally agrees well with experiments
for small- to medium-sized systems, although it may underestimate
excitation energies for long conjugated chains and charge-transfer
excitations, where range-separated hybrids such as CAM-B3LYP or higher-level
methods perform better \citep{kowalczyk2019comparative,sun2014electronic}.
B3LYP was selected as the primary TDDFT functional for detailed spectral
analysis, while HSE06, CAM-B3LYP, and PBE were used to assess functional
dependence. Next, we discuss the computed TDDFT absorption spectra
from the ground state of each molecule considered in this work. For
brevity, the discussion is restricted to the prominent peaks below
7 eV and their dominant configurations, while the SI presents a complete
analysis of all spectral peaks. 

\subsubsection*{AP1 }

The B3LYP/TDDFT (Figure ~\ref{Fig.3.}) spectrum of AP1 exhibits
four prominent absorption peaks between 3.27 and 6.32 eV involving
x and y-polarized photon absorption as described in Table S5. The
first optically allowed transition at 3.27 eV (see Table \ref{tab:peak_position})
is dominated by the $|H\rightarrow L+1\rangle$ excitation, while
the symmetry-forbidden $|H\rightarrow L\rangle$ transition is dark.
The higher-energy peaks originate from excitations involving low-lying
occupied and high-lying virtual molecular orbitals. B3LYP and HSE06
predict nearly identical spectra, whereas CAM-B3LYP blue-shifts and
PBE red-shifts the absorption peaks (see Figure S1, and Table S5 -
S8). 

\subsubsection*{AP2}

The B3LYP/TDDFT spectrum of AP2 is depicted in Figure ~\ref{Fig.4.},
and the detailed analysis of each peak is provided in Table S11. The
first two prominent features, peaks \RNum{1} and \RNum{2}, appearing
at 3.09 and 3.23 eV, originate from x- and y-polarized excitations
dominated by the $|H\rightarrow L+1\rangle_{x}$ and $|H\rightarrow L+2\rangle_{y}$
transitions, respectively. The higher-energy region consists of closely
spaced excitations of which peaks \RNum{3} and \RNum{4} arise
from mixed contributions of higher unoccupied states and lower occupied
states, with dominant transitions $|H\rightarrow L+9\rangle_{x}$,
$|H-2\rightarrow L\rangle_{x}$; and $|H-3\rightarrow L\rangle_{y}$
and $|H\rightarrow L+12\rangle_{x}$. The subsequent features (Peak
\RNum{5} to \RNum{7}) are associated with transitions to higher
virtual orbitals, exhibiting a combination of x- and y-polarized excitations. 

\subsubsection*{AP31 and AP32}

The B3LYP/TDDFT spectrum of AP31 (AP32) is presented in Figure ~\ref{Fig.5.}
(~\ref{Fig.6.}). According to Tables S14 and S17, peaks \RNum{1}
and \RNum{2} of both the systems originate from $|H\rightarrow L+1\rangle_{y}$
and $|H\rightarrow L+2\rangle_{x}$, respectively. The subsequent
peaks involve excitations from deeper occupied states to higher virtual
orbitals. For AP31, Peak III is mainly characterized by $|H-3\rightarrow L\rangle_{y}$transition,
whereas AP32 exhibits highest contributions from $|H-3\rightarrow L\rangle_{y}$
and $|H\rightarrow L+9\rangle_{x}$ excitations. The higher-energy
absorption peaks (peak \RNum{5} and beyond) in both systems consist
of multiple closely spaced transitions involving higher unoccupied
orbitals, with both x- and y-polarized contributions. 

\subsubsection*{AP41 and AP42}

The B3LYP/TDDFT spectra of AP41 (Figure ~\ref{Fig.7.}) and AP42
(Figure ~\ref{Fig.8.}) are composed entirely of mixed (xy)-polarized
electronic excitations, indicating that the transition dipole moments
have finite components along both the x- and y-directions. The lowest-energy
absorption (Peak \RNum{1}) is dominated by the $|H\rightarrow L+1\rangle_{xy}$transition
at 3.66 eV for AP41, whereas AP42 receives comparable contributions
from $|H\rightarrow L+1\rangle_{xy}$ (3.79 eV), $|H\rightarrow L+2\rangle_{xy}$
(3.91 eV), respectively. Peak \RNum{2} arises primarily from $|H\rightarrow L+2\rangle_{xy}$
in AP41, while in AP42 it is governed by the $|H-3\rightarrow L\rangle_{xy}$
(5.49 eV) and $|H-4\rightarrow L\rangle_{xy}$ (5.54 eV) transitions.
The higher-energy region of the AP41 spectrum is dominated by excitations
from lower-lying occupied molecular orbitals to the LUMO and nearby
virtual orbitals, while the corresponding spectral region in AP42
contains only a few low-intensity transitions. Detailed excited-state
assignments are listed in Tables S20 and S23.

\subsubsection*{AP5}

As shown in Figure ~\ref{Fig.9.}, peaks \RNum{1} (3.79 eV) and
\RNum{2} (4.06 eV) are mainly dictated by $|H\rightarrow L+1\rangle_{x}$
and $|H\rightarrow L+2\rangle_{y}$, transitions, respectively. Peak
\RNum{3} arises mainly from the $|H-4\rightarrow L\rangle_{y}$
and $|H-3\rightarrow L\rangle_{x}$ configurations from excitation
at 5.35 eV and 5.40 eV, respectively, while peaks \RNum{4} to \RNum{6}
are governed by transitions involving deeper occupied and higher virtual
orbitals. The complete excited-state assignments for this molecule
are listed in Table S26.

\subsubsection*{AP61}

Peak \RNum{1} for AP61 derives its main contribution from the excitation
$|H\rightarrow L+1\rangle_{xy}$ (4.19 eV), as depicted in Figure
~\ref{Fig.10.}, while the main contribution to peak \RNum{2} comes
from $|H\rightarrow L+2\rangle_{xy}$ (4.39 eV). According to Table
S29, peaks \RNum{3}, \RNum{4}, and \RNum{5} derive their main
contributions from the excitations $|H-4\rightarrow L\rangle_{xy}$
(5.69 eV), $|H-7\rightarrow L\rangle_{xy}$ (5.98 eV), and $|H-4\rightarrow L+1\rangle_{xy}$
(6.30 eV), respectively. 

\subsubsection*{AP7}

The B3LYP/TDDFT absorption spectrum of AP7 is shown in Figure ~\ref{Fig.11.},
with the corresponding excitation details summarized in Table S32.
The first absorption peak at 4.34 eV arises from nearly degenerate
$|H\rightarrow(L+1)_{2}\rangle_{x}$ and $|H\rightarrow(L+1)_{1}\rangle_{y}$
transitions involving both x- and y-polarized photons, while the second
peak at 5.75 eV is dominated by transitions from the $|(H-5)_{1}\rightarrow L\rangle_{x}$
and $|(H-5)_{2}\rightarrow L\rangle_{y}$, reflecting the contribution
of deeper valence states to the higher-energy optical response.

In conclusion, the TDDFT spectra demonstrate that structural modifications
alter the optical response by changing the symmetry and spatial character
of the excited states. This variation in polarization-dependent transitions
and the involvement of deeper occupied and higher virtual molecular
orbitals indicate that the absorption characteristics are governed
by the detailed electronic structure of each framework, moving far
beyond a simple frontier orbital picture.

\begin{table}[th]
\centering
\caption{First prominent absorption peak positions in the ground-state spectra
of N-substituted phenalenyl molecules }\label{tab:peak_position}
\smallskip{}
\begin{tabular}{lccccc}
\toprule 
\multirow{3}{*}{\textbf{System}} & \multicolumn{5}{c}{\textbf{Peak position (in eV)}}\tabularnewline
\cmidrule{2-6}
 & \multicolumn{1}{c}{PPP-FCI} & \multicolumn{4}{c}{TDDFT}\tabularnewline
\cmidrule{2-6}
 & \multicolumn{1}{c}{OHNO-\RNum{3}} & \multirow{1}{*}{B3LYP} & \multirow{1}{*}{HSE06} & \multirow{1}{*}{CAM-B3LYP} & \multirow{1}{*}{PBE}\tabularnewline
\midrule 
AP1 & 2.65 & 3.27 & 3.33 & 3.49 & 3.08\tabularnewline
AP2 & 2.47 & 3.09 & 3.14 & 3.28 & 2.92\tabularnewline
AP31 & 2.94 & 3.52 & 3.60 & 3.80 & 3.27\tabularnewline
AP32 & 2.31 & 3.06 & 3.11 & 3.23 & 2.90\tabularnewline
AP41 & 3.25 & 3.66 & 3.73 & 3.96 & 3.39\tabularnewline
AP42 & 3.30 & 3.79 & 3.86 & 4.09 & 3.51\tabularnewline
AP5 & 3.58 & 3.79 & 3.86 & 4.09 & 3.50\tabularnewline
AP61 & 3.76 & 4.19 & 4.28 & 4.54 & 3.84\tabularnewline
AP7 & 4.12 & 4.34 & 4.43 & 4.70 & 3.97\tabularnewline
\bottomrule
\end{tabular}
\end{table}

\subsubsection{PPP-FCI Spectra}

Given that we have already discussed $S_{0}\rightarrow S_{1}$ transitions
in azaphenalenes previously, in this section, we present their calculated
PPP-FCI optical absorption spectra from the ground state $S_{0}$,
to the higher singlet states $S_{n}$( $n\geq2$). For all the molecules,
the spectra were computed using the PPP-FCI method, employing the
OHNO-\RNum{3} parametrization, and the complete analysis of many-particle
wave functions of the excited states, giving rise to peaks in the
computed spectra, is provided in the SI. In these calculations, we
employed the point-group symmetry $C_{s}$ to reduce the computational
time. However, to ensure the correctness of the symmetry-based results,
we also performed a few calculations using no symmetry ($C_{1}$),
and obtained identical results. Therefore, we are confident of the
accuracy of our results presented next. For all the molecules, the
dominant wave functions of the $S_{0}$ and $S_{1}$ states are provided
Table S35.

\subsubsection*{AP1}

Peak \RNum{1} of the FCI spectrum of AP1 is located at 2.65 eV in
Fig. ~\ref{Fig.3.}, and it is composed of two degenerate peaks to
the excited states of A$^{\prime}$ ($y$-polarized) and A$^{\prime\prime}$
($x$-polarized) irreducible representations (see Table S9), corresponding
to the $S_{0}\rightarrow S_{2}$ transition. The A$^{\prime}$ peak
mainly arises from the $|H\rightarrow(L+1)_{2}\rangle_{y}$ configuration,
and A$^{\prime\prime}$ peak is from $|H\rightarrow(L+1)_{1}\rangle_{x}$.
On the other hand, the first B3LYP/TDDFT peak at 3.27 eV (see Table
\ref{tab:peak_position}). Therefore, FCI predicts the first prominent
optically allowed absorption feature in a much lower energy region
because of its superior inclusion of the electron correlation effects,
which generate additional low-lying excited states through collective
determinant mixing. On the other hand, conventional adiabatic TDDFT
may not adequately describe excited states with substantial multiconfigurational
character and, consequently, may miss some correlation-driven spectral
features, resulting in fewer predicted peaks \citep{harrison1991approximating,illas1991selected}.
Peak \RNum{2} consists of two degenerate peaks at 4.03 eV, which
derive their dominant contributions from the doubly-excited configurations
$|H\rightarrow L;H\rightarrow(L+1)_{1}\rangle_{y}$ (A$^{\prime}$)
and $|H\rightarrow L;H\rightarrow(L+1)_{2}\rangle_{x}$ (A$^{\prime\prime}$).
Peak \RNum{3}, the highest intensity peak at 5.91 eV is dominated
by single excitations $|H-2\rightarrow(L+1)_{1}\rangle_{y}$ and $|H-2\rightarrow(L+1)_{2}\rangle_{x}$.
A close comparison of the FCI (Table S9) and TDDFT (Table S5) spectra
shows that while the same configurations appear in both, their coefficients
differ, and FCI wave functions additionally contain significantly
higher-order excitations. 

\subsubsection*{AP2}

The $S_{0}\rightarrow S_{2}$ transition, composed of two degenerate
peaks corresponding to the excited states of A$^{\prime}$ and A$^{\prime\prime}$symmetries
(see Table S12), represents the first optically allowed transition,
and gives rise to peak \RNum{1} in the absorption spectrum of AP2
(see Figure ~\ref{Fig.4.}). This peak, located at 2.47 eV (Table
\ref{tab:peak_position}), is characterized by wave functions dominated
by the single excitations $|H\rightarrow L+2\rangle_{y}$ (A$^{\prime}$)
and $|H\rightarrow L+1\rangle_{x}$ (A$^{\prime\prime}$). Peaks \RNum{2}
(3.72 eV) and \RNum{3} (3.87 eV) of the PPP-FCI spectrum are very
close, and are dominated by $|H\rightarrow L;H\rightarrow L+1\rangle_{y}$
and $|H\rightarrow L;H\rightarrow L+2\rangle_{x}$ double excitations,
respectively. The next peak (\RNum{4}) mainly arises from the configurations
$|H-1\rightarrow L\rangle_{x}$ (4.25 eV) and $|H\rightarrow L+1;H\rightarrow L+1\rangle_{y}$
(4.32 eV), while peak \RNum{5} is primarily composed of excitations
$|H-2\rightarrow L\rangle_{y}$ and $|H-1\rightarrow L\rangle_{x}$.
Peak \RNum{6} consists of three closely spaced transitions dominated
by $|H\rightarrow L;H-3\rightarrow L\rangle_{x}$(5.74 eV), $|H-2\rightarrow L+2\rangle_{x}$(5.82
eV), and $|H-1\rightarrow L+2\rangle_{y}$(5.82 eV). When we compare
the computed PPP-FCI and TDDFT optical gaps, we note that all the
functionals predict larger values, with the PBE result being closest
to the PPP-FCI value.

\subsubsection*{AP31 and AP32}

In AP31 (see Fig. \ref{Fig.5.}), peak \RNum{1} (2.94 eV) is dominated
by single excitation $|H\rightarrow L+1\rangle_{y}$, while peak \RNum{2}
(3.22 eV) arises primarily from $|H\rightarrow L+2\rangle_{x}$. The
prominent peak (\RNum{5}) splits into two nearly degenerate components
at 4.59 eV (A$^{\prime\prime}$) and 4.62 eV (A$^{\prime}$), with
the highest contributions from excitations $|H-2\rightarrow L\rangle_{x}$,
and $|H\rightarrow L;H\rightarrow L\rangle_{y}$, respectively. For
AP32 (see Figure ~\ref{Fig.6.}), peak \RNum{1}consists of two
closely spaced transitions at 2.31 and 2.34 eV, dominated by double
excitations $|H\rightarrow L;H\rightarrow L+2\rangle_{x}$ and $|H\rightarrow L;H\rightarrow L+1\rangle_{y}$
, respectively. Similarly, peak \RNum{2} comprises transitions at
3.57 and 3.60 eV, with dominant contributions from $|H\rightarrow L+1\rangle_{x}$
and $|H\rightarrow L+2\rangle_{y}$, respectively. A complete analysis
of the PPP-FCI absorption spectra is provided in Table S15 (AP31)
and S18 (AP32). 

\subsubsection*{AP41 and AP42}

In the FCI spectra of AP41 and AP42 (Figs. ~\ref{Fig.7.} and ~\ref{Fig.8.}),
peak \RNum{1} appears at 3.25 and 3.30 eV, respectively, and is mainly
governed by the $|H\rightarrow L+1\rangle_{xy}$ configuration. The
corresponding TDDFT peaks occur at 3.66 eV (AP41) and 3.79 eV (AP42).
In the two molecules, peak \RNum{2} is located at 3.39 (AP41) and
3.41 eV (AP42), and in both cases is dominated by the $|H\rightarrow L+2\rangle_{xy}$
configuration (see Tables S21, and S24). 

\subsubsection*{AP5}

Peak \RNum{1} of AP5 (Figure ~\ref{Fig.9.}) splits into two closely
spaced transitions at 3.58 and 3.65 eV, which derive their main contributions
from $|H\rightarrow L+2\rangle_{x}$configuration under A$^{\prime\prime}$symmetry,
and $|H\rightarrow L+1\rangle_{y}$ configuration under A$^{\prime}$symmetry,
respectively (see Table S27). The corresponding TDDFT peak is predicted
at 3.79 eV (given in Table S25). Peak \RNum{2} is mainly dominated
by the double excitation $|H\rightarrow L;H\rightarrow L+2\rangle_{y}$(4.73
eV) and the single excitation $|H-1\rightarrow L\rangle_{x}$ (4.85
eV).

\subsubsection*{AP61}

In the absorption spectrum of AP61, the first FCI peak is composed
of two subpeaks at 3.76 eV and 3.89 eV, to be compared with the first
TDDFT peak at 4.19 eV. The FCI subpeak at 3.76 eV is dominated by
the configuration $|H\rightarrow L+1\rangle_{xy}$, while the one
at 3.89 eV is mainly composed of $|H\rightarrow L+2\rangle_{xy}$
excitation. The next peak appears at 4.98 eV with the highest contribution
from $|H-1\rightarrow L\rangle_{xy}$. A detailed configuration analysis
of the ground-state absorption spectrum of AP61 is provided in Table
S30.

\subsubsection*{AP7}

For AP7, the FCI peak \RNum{1} appears at 4.12 eV (4.34 eV for TDDFT)
corresponding to the doubly degenerate $S_{0}\rightarrow S_{2}$ transition
(see Figure ~\ref{Fig.11.}). Dominant configuration for this peak
are $|H\rightarrow(L+1)_{2}\rangle_{y}$ and $|H\rightarrow(L+1)_{1}\rangle_{x}$
with A$^{\prime}$, and A$^{\prime\prime}$ symmetries, respectively.
The most intense absorption at 6.38 eV is also doubly degenerate and
their wave functions exhibit significant configuration mixing with
several configurations of almost equal coefficients. The configurations
with the largest coefficients are $|(H-2)\rightarrow(L+1)_{1}\rangle_{y}$
(A$^{\prime}$) and $|(H-2)\rightarrow(L+1)_{2}\rangle_{x}$ (A$^{\prime\prime}$),
respectively. There is another sharp transition at 6.64 eV, with the
dominant contribution from the same configurations. Detailed analysis
of all spectral peaks of AP7 is presented in Table S33.

\subsection{Triplet Optical Absorption}

Using the PPP-FCI method, we also calculated the excited state absorption
spectra of azaphenalene molecules from their $T_{1}$ states to higher
triplet states ($T_{1}\rightarrow T_{n},\:n\geq2$). Although these
systems exhibit singlet--triplet inversion ($S_{1}<T_{1}$), the
triplet excited-state manifold ($T_{1}$, $T_{2}$, ...) remains well
defined, and supports spin-allowed $T_{1}\rightarrow T_{n}$ transitions.
Furthermore, a study of $T_{1}\rightarrow T_{n}$ transitions provides
insights into the nature of triplet states of these molecules. 

\subsubsection*{AP1}

\begin{figure}[H]
\centering\includegraphics[width=6.5in,totalheight=5in]{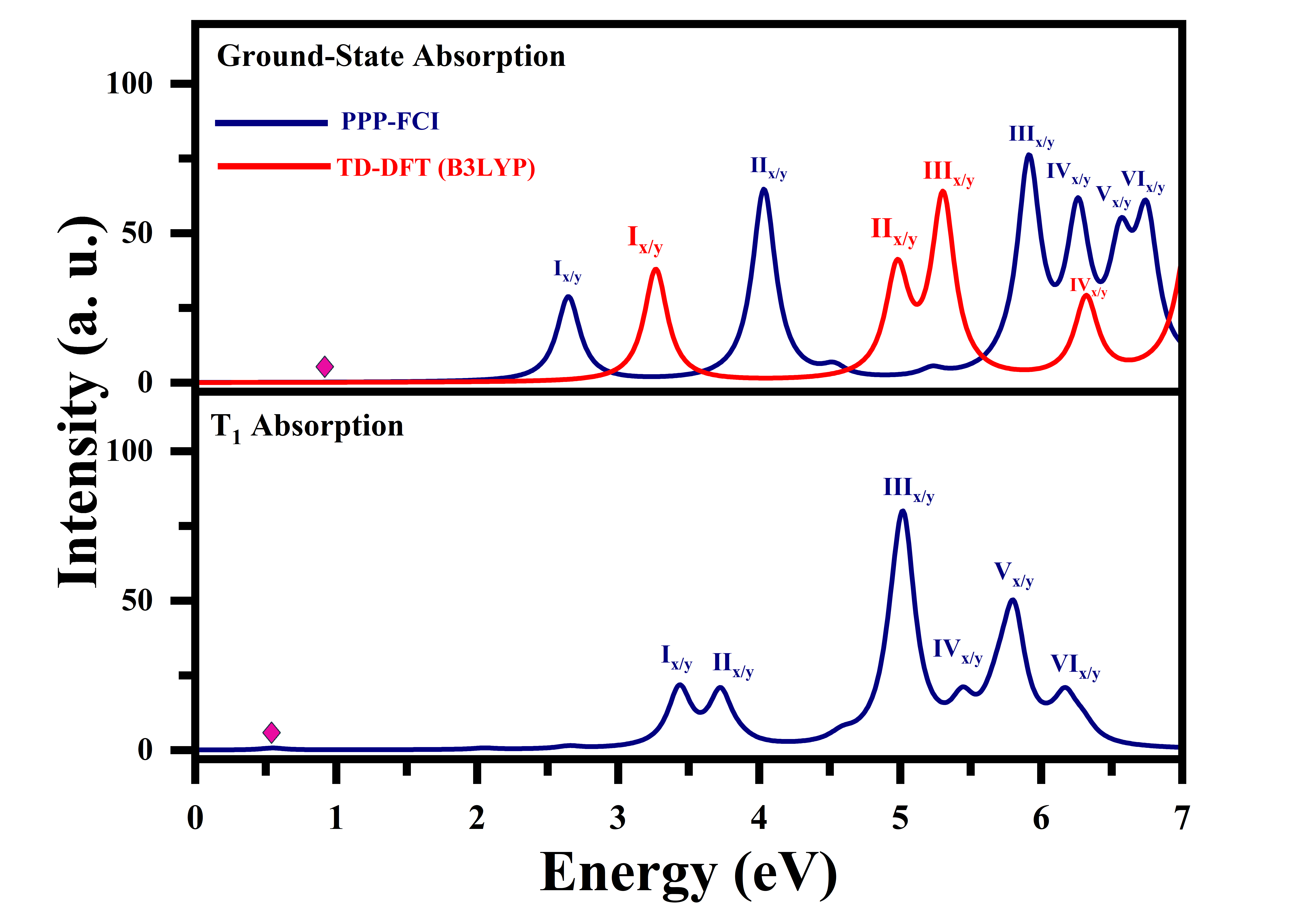}\caption{For AP1, i.e., cycl{[}3.3.3{]}azine, the ground-state absorption spectra
(top-panel) calculated using PPP-FCI (OHNO-\RNum{3}) and TDDFT methods,
and the triplet absorption spectrum from the $T_{1}$ state (bottom-panel)
calculated using the PPP-FCI (OHNO-\RNum{3}) method. The subscripts
of the peak labels denote polarization directions of each peak, while
pink diamond markers are used to denote the locations of locations
of the $S_{0}\rightarrow S_{1}$ and $T_{1}\rightarrow T_{2}$ absorptions.
}\label{Fig.3.}
\end{figure}

In the $T_{1}$ absorption spectrum of AP1 (Fig. \ref{Fig.3.}), the
$T_{1}\rightarrow T_{2}$ transition is weakly allowed, with an oscillator
strength of $f=0.0029$ at 0.55 eV, and the $T_{2}$ wave function
is mainly composed of the configuration $|H\rightarrow(L+1)_{2}\rangle$,
accompanied by minor contributions from $|H\rightarrow(L+1)_{1};H\rightarrow(L+1)_{2}\rangle$,
$|H\rightarrow L;(H-1)_{1}\rightarrow(L+1)_{1}\rangle+c.c$, and $|H\rightarrow L;H-2\rightarrow L\rangle$.
The first intense absorption (Peak \RNum{1}) appears at 3.43 eV
(Table S10), and is composed of two degenerate excited states whose
wave functions are dominated by the configurations $|(H-1)_{1}\rightarrow L\rangle_{x}$
and $|(H-1)_{2}\rightarrow L\rangle_{y}$. The information about the
rest of the peaks is presented in Table S10.

\subsubsection*{AP2}

\begin{figure}
\centering\includegraphics[width=6.5in,totalheight=5in]{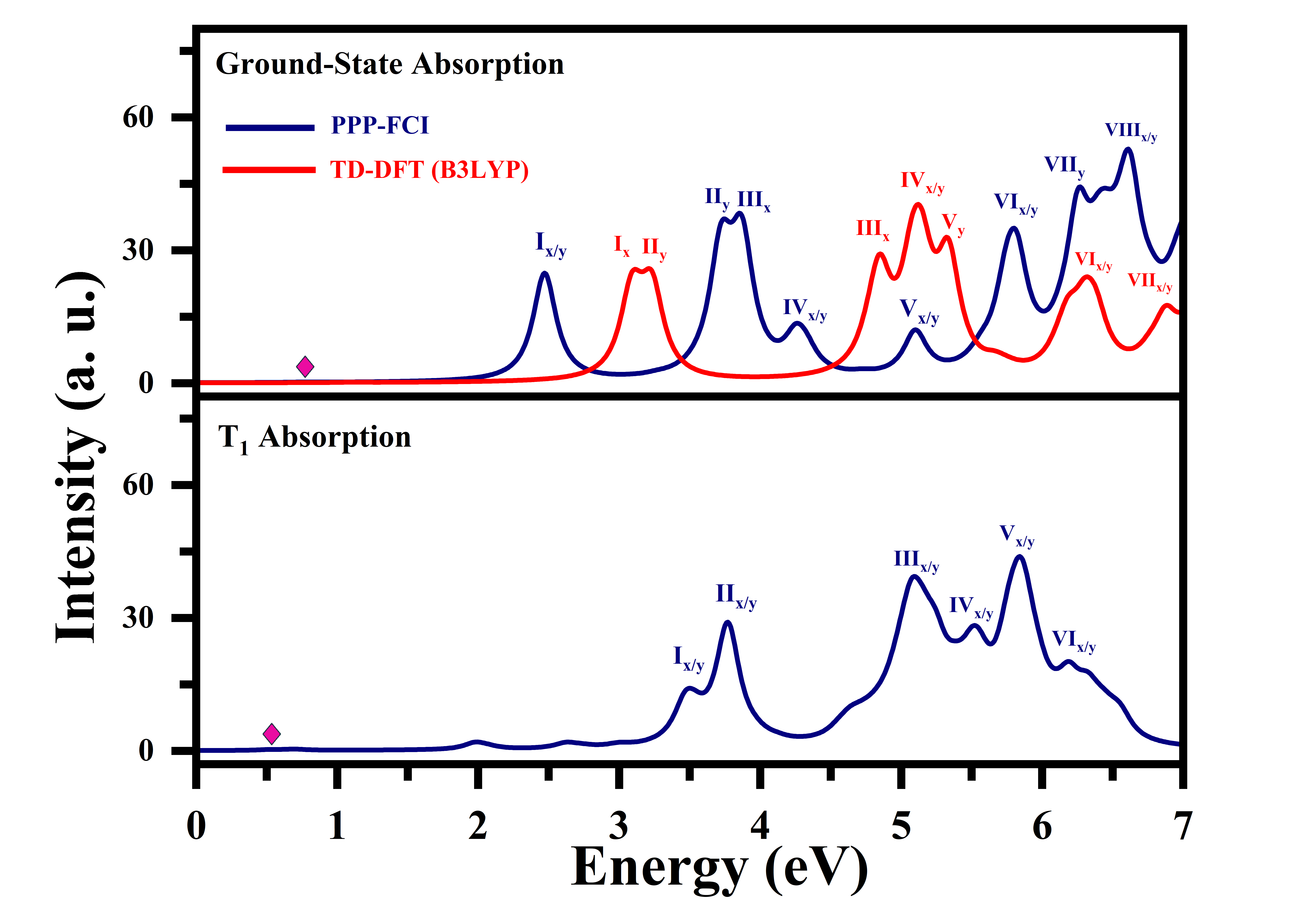}

\caption{The ground-state and triplet absorption spectra of AP2, i.e., 2-monoazacycl{[}3.3.3{]}azine.
The rest of the information is same as that in the caption of Fig.
\ref{Fig.3.}. }\label{Fig.4.}
\end{figure}

In AP2, the $T_{1}\rightarrow T_{2}$ transition is weakly allowed
at 0.52 eV with an oscillator strength of $f=0.0015$, and the $T_{2}$
wave function is dominated by $|H\rightarrow L+1\rangle,$with small
contribution from $|H\rightarrow L+1;H-1\rightarrow L+2\rangle$ and
$|H\rightarrow L;H\rightarrow L+2\rangle$. The first intense absorption,
peak \RNum{1} (Figure ~\ref{Fig.4.}) arises from three closely
spaced transitions to the states of A$^{\prime}$ and A$^{\prime\prime}$
symmetries, corresponding to the absorption of x- and y-polarized
photons, respectively (see Table S13). The excited states giving rise
to peak \RNum{1} are dominated by configurations $|H\rightarrow L+5\rangle_{y}$
(3.46 eV), $|H\rightarrow L+5\rangle_{y}$(3.47 eV), and $|H-1\rightarrow L+2\rangle_{x}$
(3.54 eV). Peak \RNum{2} of AP2 is composed of two almost overlapping
transitions corresponding to the absorptions of x- and y-polarized
photons to the excited states dominated by configurations $|H-3\rightarrow L\rangle_{x}$
(3.76 eV), and $|H-1\rightarrow L\rangle_{y}$ (3.78 eV), respectively.
The detailed information about the $T_{1}$ absorption spectrum of
AP2 is presented in Table S13.

\subsubsection*{AP31 and AP32}

\begin{figure}[!t]
\centering \includegraphics[width=6.5in,totalheight=5in,height=5.5in]{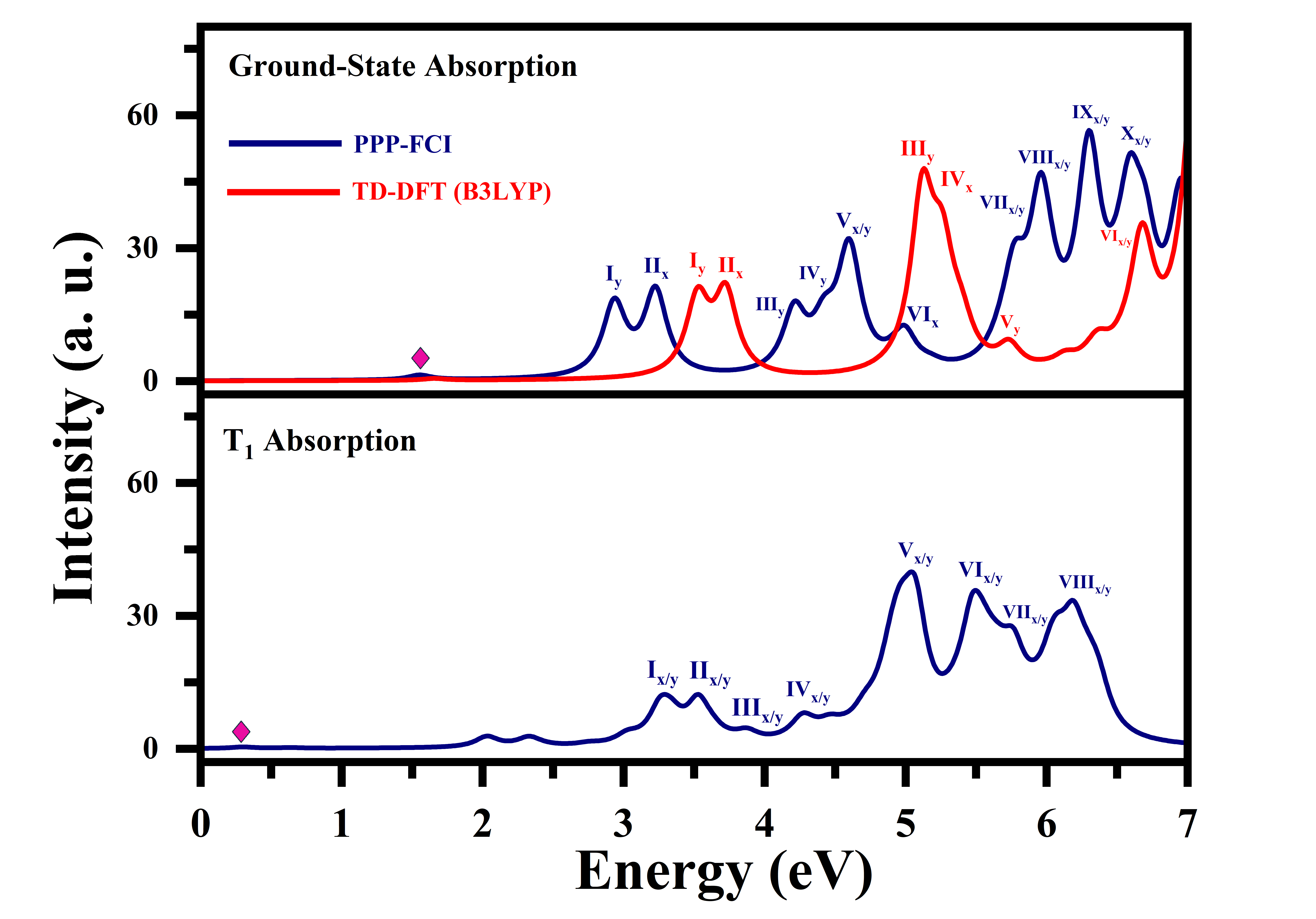}
\caption{The ground-state and triplet absorption spectra of AP31, i.e., 6,7-diazacycl{[}3.3.3{]}azine.
The rest of the information is same as that in the caption of Fig.
\ref{Fig.3.}.  }\label{Fig.5.}
\end{figure}

In AP31, the weakly allowed $T_{1}\rightarrow T_{2}$ transition with
$f=0.0027$ appears at 0.30 eV (Figure ~\ref{Fig.5.}), and the $T_{2}$
wave function is dominated by $|H\rightarrow L+1\rangle$ configuration,
with other significant contributions from double excitations $|H\rightarrow L+1;H-1\rightarrow L+2\rangle$
and $|H\rightarrow L+1;H-2\rightarrow L+1\rangle$. Peak \RNum{1}
of AP31 comprises a doubly degenerate transition at 3.26 eV, and a
transition at 3.34 eV (see Table S16). The degenerate transitions
are dominated by $|H-2\rightarrow L+1\rangle_{x}$ and $|H\rightarrow L+5\rangle_{y}$,
while the transition at 3.34 eV is dominated by the double excitation
$|H\rightarrow L;H\rightarrow L+1\rangle_{y}$ . Details of the other
peaks of $T_{1}$ absorption spectrum are provided in Table S16.

\begin{figure}[!t]
\centering \includegraphics[width=6.5in,totalheight=5in,height=5.5in]{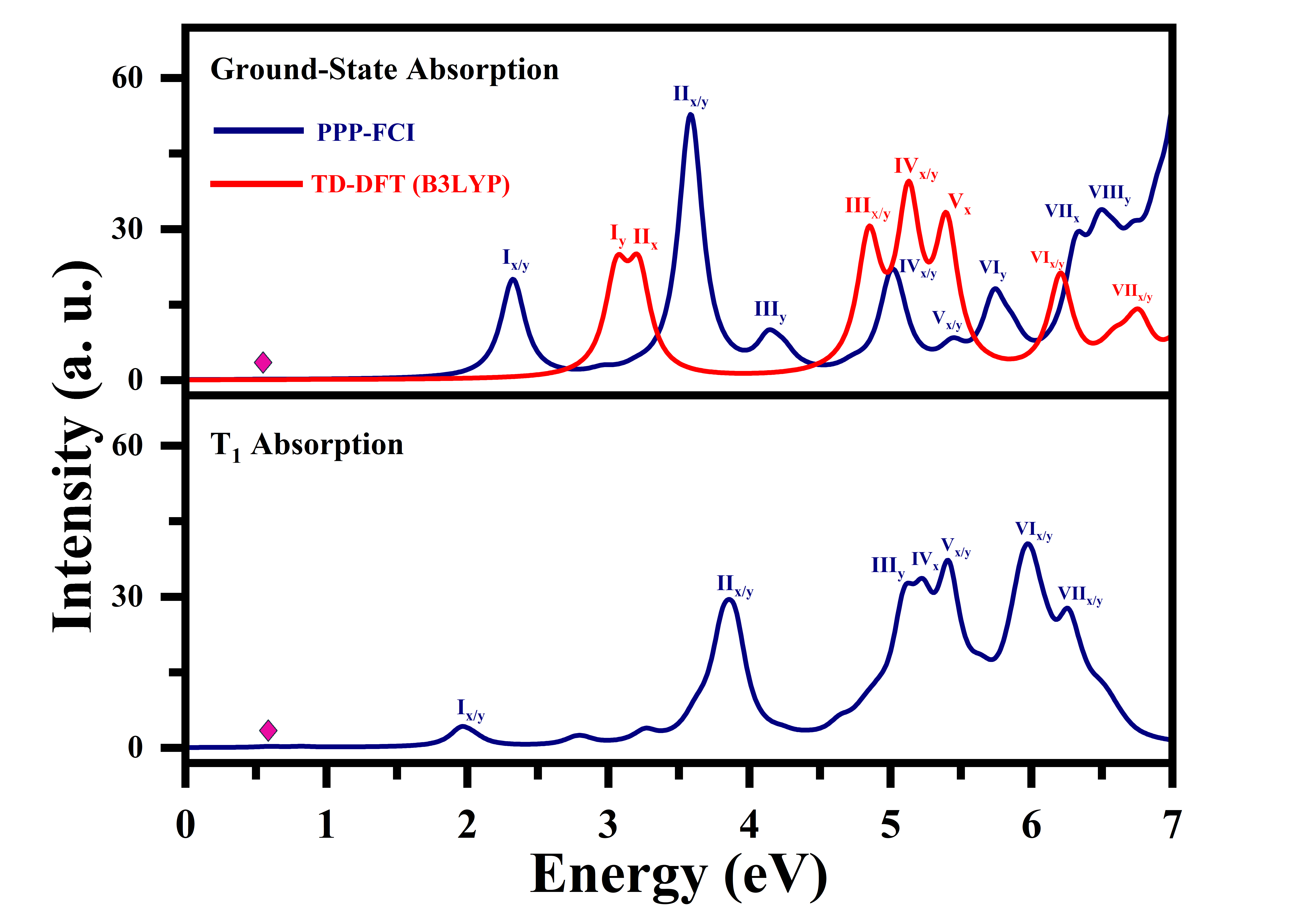}
\caption{The ground-state and triplet absorption spectra of AP32, i.e., 5,8-diazacycl{[}3.3.3{]}azine.
The rest of the information is same as that in the caption of Fig.
\ref{Fig.3.}.  }\label{Fig.6.}
\end{figure}

The $T_{1}$ FCI spectrum of AP32 reveals that, $T_{1}\rightarrow T_{2}$
transition with $f=0.0015$ appears at 0.59 eV (see Fig. ~\ref{Fig.6.}),
with the highest contribution in the $T_{2}$ wave function is from
the double excitation $|H\rightarrow L;H\rightarrow L+1\rangle$,
and smaller contributions from $|H\rightarrow L;H\rightarrow L+3\rangle$,
$|H\rightarrow L+2\rangle$, $|H\rightarrow L;H-3\rightarrow L+1\rangle$,
and $|H\rightarrow L;H\rightarrow L+1;H-1\rightarrow L+2\rangle$,
which is a triple excitation. The first intense feature of the AP32
spectrum is dominated by $|H\rightarrow L+1\rangle_{y}$ (at 1.96
eV) and $|H\rightarrow L+2\rangle_{x}$(at 2.05 eV). Peak \RNum{2}
of AP32 is due to the excited states at 3.81 eV and 3.90 eV, with
dominant contributions from the configurations $|H\rightarrow L;H-1\rightarrow L\rangle_{x}$
and $|H\rightarrow L;H-2\rightarrow L\rangle_{y}$, respectively (Table
S19). The corresponding data for the remaining peaks are listed in
Table S19.

\subsubsection*{AP41 and AP42}

\begin{figure}[!t]
\centering \includegraphics[width=6.5in,totalheight=5in]{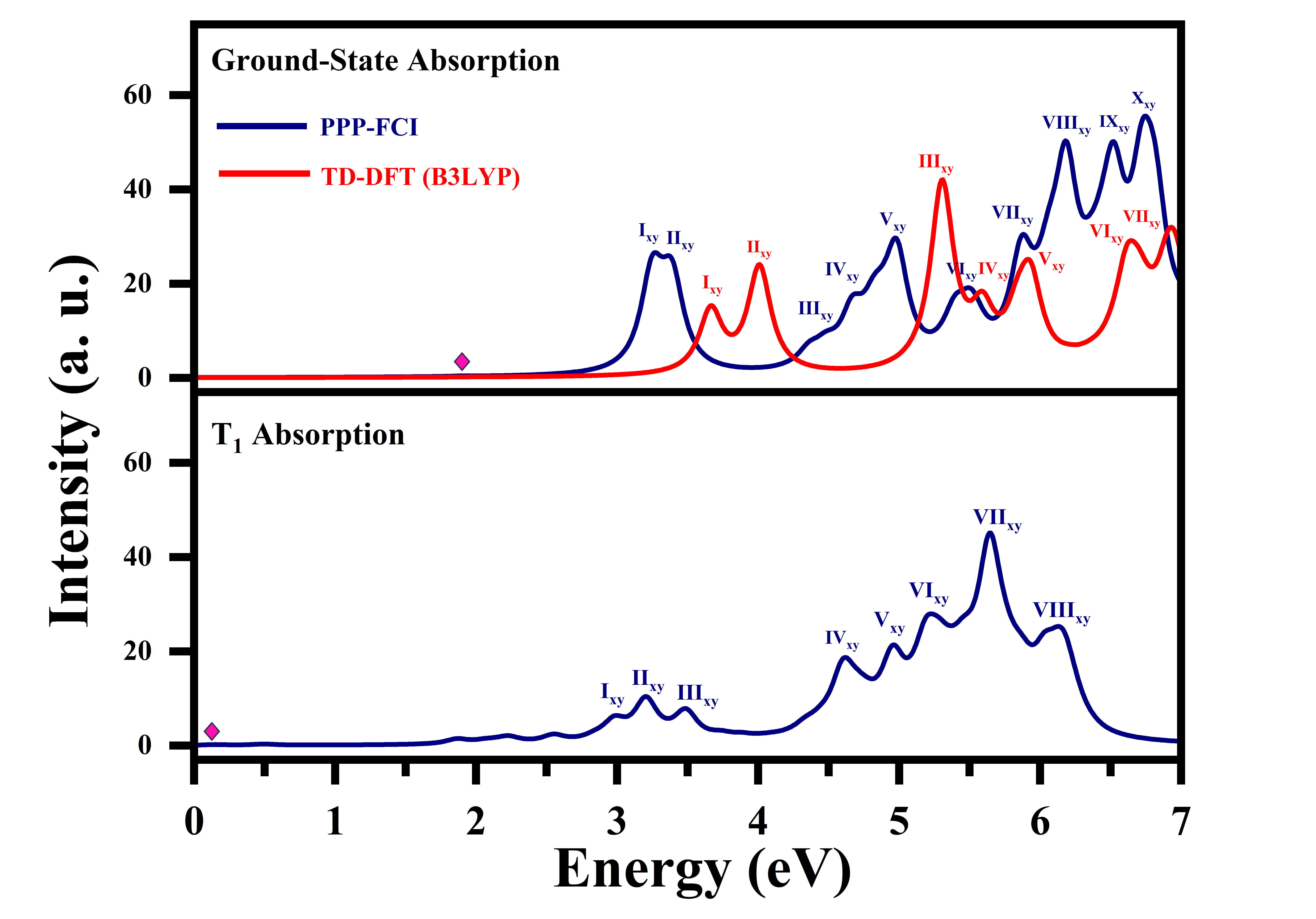}
\caption{The ground-state and triplet absorption spectra of AP41, i.e., 1,3,6-triazacycl{[}3.3.3{]}azine.
The rest of the information is same as that in the caption of Fig.
\ref{Fig.3.}.  }\label{Fig.7.}
\end{figure}

\begin{figure}[!t]
\centering \includegraphics[width=6.5in,totalheight=5in]{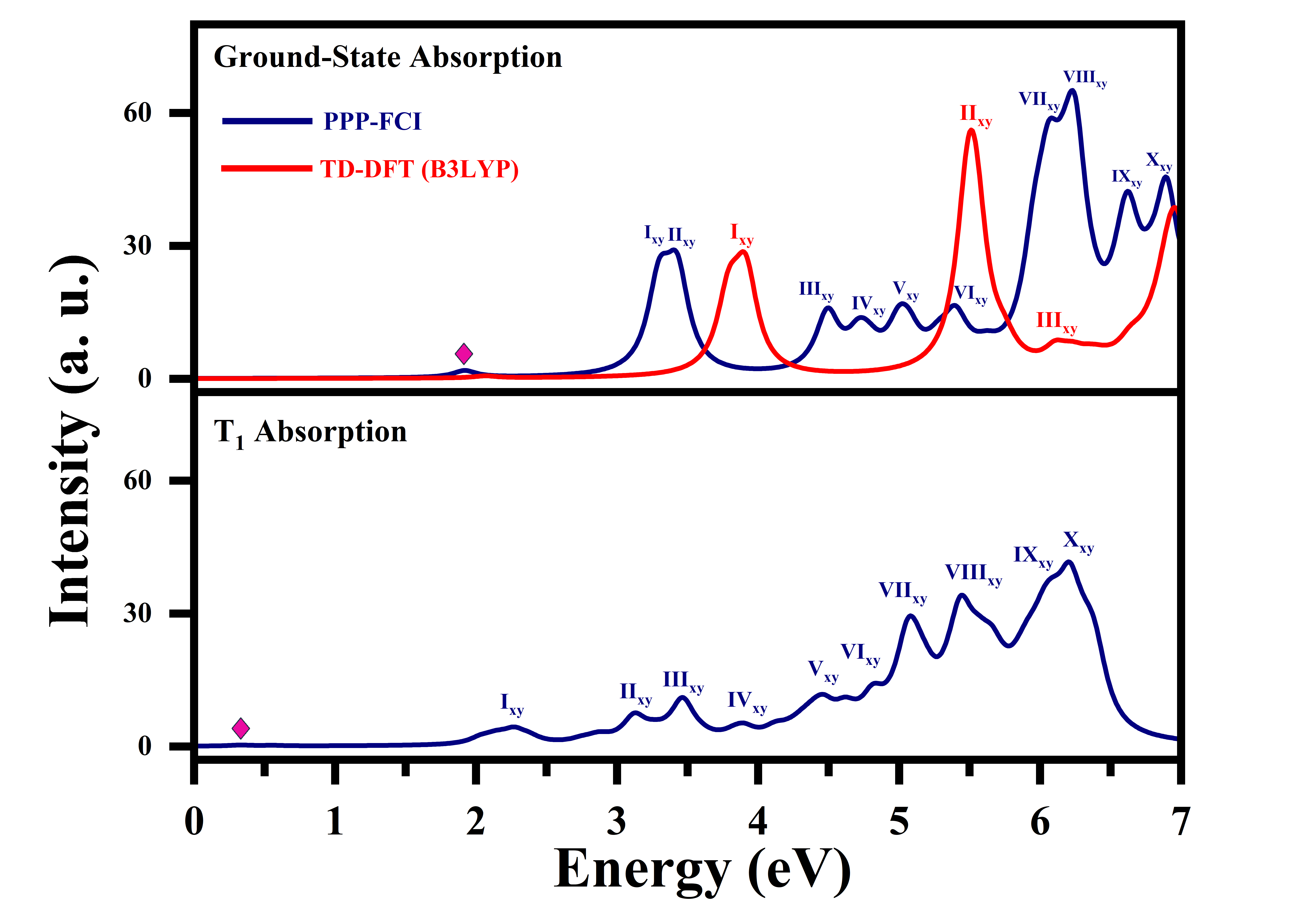}
\caption{The ground-state and triplet absorption spectra of AP42, i.e., 1,3,4-triazacycl{[}3.3.3{]}azine.
The rest of the information is same as that in the caption of Fig.
\ref{Fig.3.}.  }\label{Fig.8.}
\end{figure}

For AP41 and AP42, the $T_{1}\rightarrow T_{2}$ transitions appear
at 0.15 eV ($f=0.0012$) and 0.33 eV ($f=0.0019$), respectively (Figures
~\ref{Fig.7.} and \ref{Fig.8.}). $T_{2}$ wave function of AP41
is dominated by $|H\rightarrow L+1\rangle$, with a significant contribution
also from $|H\rightarrow L\rangle$, along with minor contributions
from the configurations $|H\rightarrow L;H-2\rightarrow L+1\rangle$,
$|H\rightarrow L+2\rangle$, and $|H\rightarrow L+1;H\rightarrow L+2\rangle$.
On the other hand, $T_{2}$ wave function of AP42 is dominated by
$|H\rightarrow L+1\rangle$, followed by configurations $|H\rightarrow L+2\rangle$,
$|H\rightarrow L+1;H-2\rightarrow L+1\rangle$, and $|H\rightarrow L+1;H-1\rightarrow L+2\rangle$.
Peak \RNum{1} of AP41 at 2.98 eV is dominated by $|H-2\rightarrow L+2\rangle_{xy}$(see
Table S22), while that for AP42 at 2.27 eV is dominated by $|H\rightarrow L+4\rangle_{xy}$
(see Table S25). Additional peak information for AP41 (AP42) is compiled
in Table S22 (Table S25).

\subsubsection*{AP5}

\begin{figure}[!t]
\centering \includegraphics[width=6.5in,totalheight=5in]{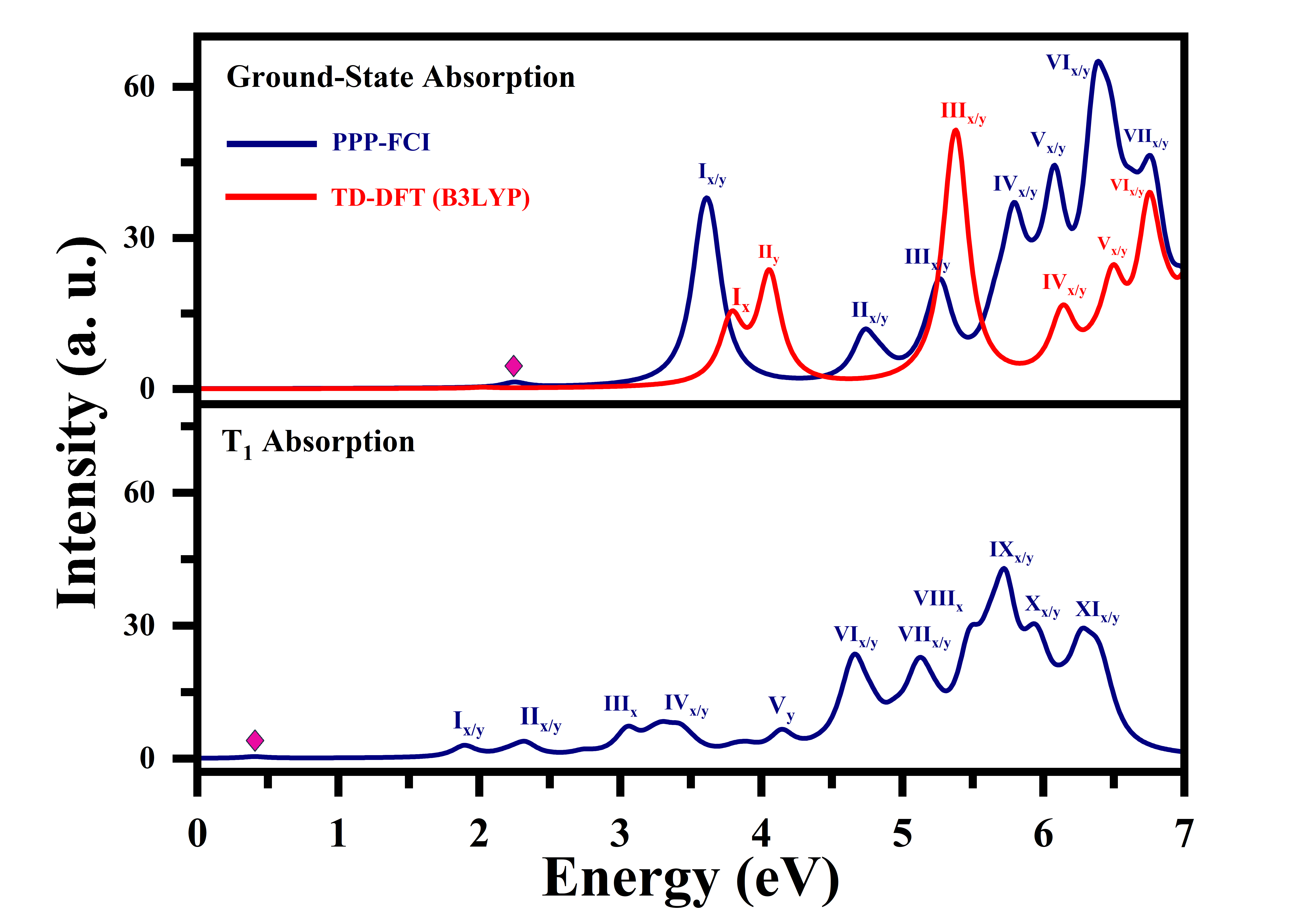}
\caption{The ground-state and triplet absorption spectra of AP5, i.e., 1,3,4,-tetraazacycl{[}3.3.3{]}azine.
The rest of the information is same as that in the caption of Fig.
\ref{Fig.3.}.  }\label{Fig.9.}
\end{figure}

For AP5, $T_{1}\rightarrow T_{2}$ transition appears at 0.40 eV with
$f=0.0026$, and its $T_{2}$ wave function derives primary contributions
from the single excitation $|H\rightarrow L+1\rangle$, along with
small contributions from $|H\rightarrow L+1;H-3\rightarrow L+2\rangle$,
$|H\rightarrow L;H-2\rightarrow L+1\rangle$, $|H-2\rightarrow L+3\rangle$,
and $|H\rightarrow L+1;H-1\rightarrow L+1\rangle$. Peak \RNum{1}
(see Fig. ~\ref{Fig.9.}) consists of two closely located transitions
at 1.89 and 1.93 eV, dominated by $|H-2\rightarrow L\rangle_{x}$
and $|H-1\rightarrow L\rangle_{y}$ , respectively (see Table S28).
The analysis of the remaining spectral features are listed in Table
S28.

\subsubsection*{AP61}

\begin{figure}[!t]
\centering \includegraphics[width=6.5in,totalheight=5in]{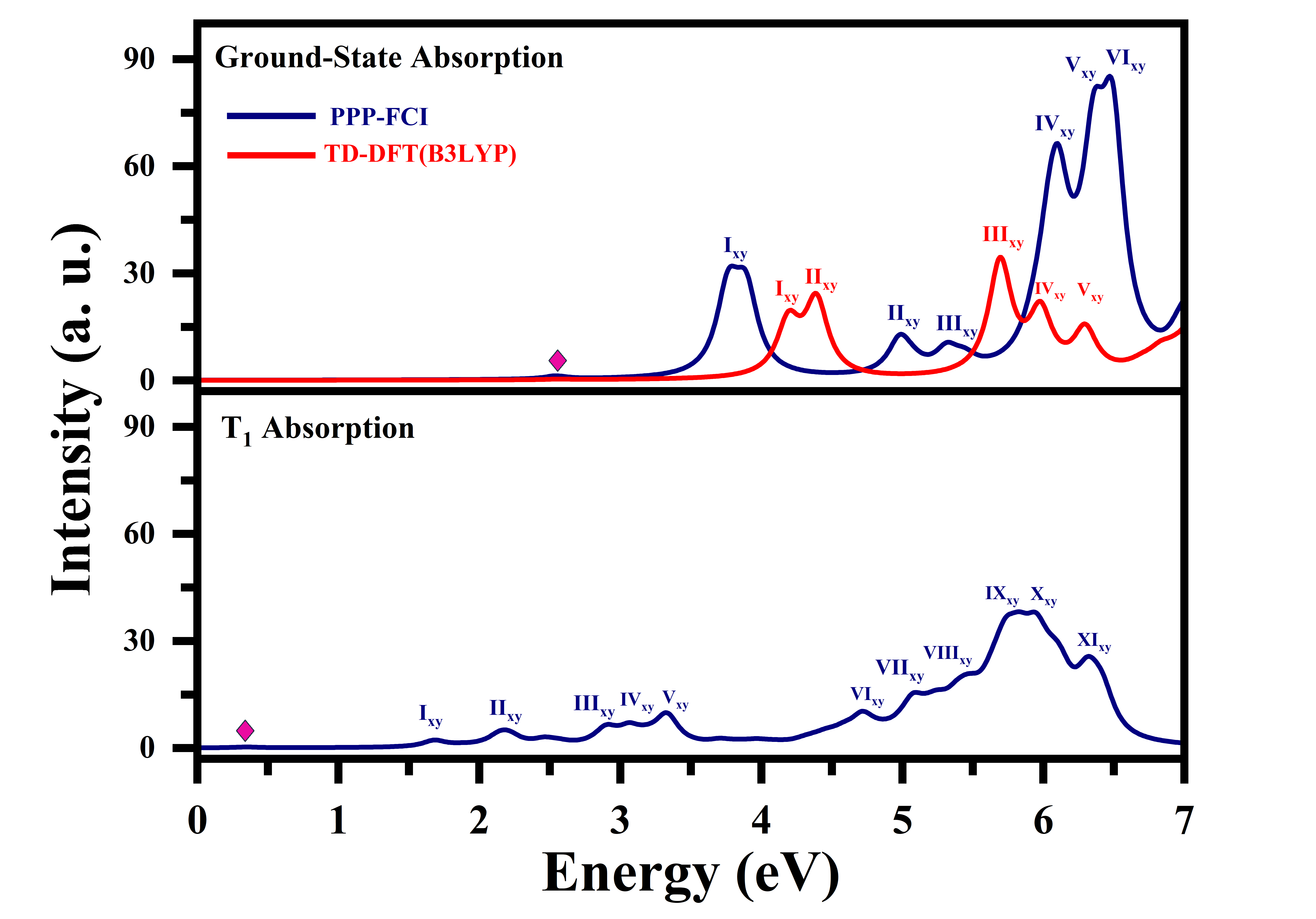}
\caption{The ground-state and triplet absorption spectra of AP61, i.e., 1,3,4,6,7-pentaazacycl{[}3.3.3{]}azine.
The rest of the information is same as that in the caption of Fig.
\ref{Fig.3.}.  }\label{Fig.10.}
\end{figure}

For AP61, $T_{1}\rightarrow T_{2}$ transition appears at 0.33 eV,
with $f=0.0011$ (see Figure ~\ref{Fig.10.}), and the $T_{2}$ wave
function mainly consists of the configurations $|H\rightarrow L+2\rangle$
and $|H\rightarrow L+1\rangle$, along with secondary contributions
from $|H-1\rightarrow L+1\rangle$, $|H\rightarrow L;H-2\rightarrow L+2\rangle$,
$|H-1\rightarrow L+2\rangle$, $|H\rightarrow L+2;H-1\rightarrow L+2\rangle$,
$|H-2\rightarrow L+4\rangle$, and $|H-1\rightarrow L+3\rangle$.
Peak \RNum{1} of AP61 is located at 1.68 eV, and the dominant configuration
contributing to it is $|H-1\rightarrow L\rangle_{xy}$ (see Table
S31). Details of the other peaks are provided in Table S31.

\subsubsection*{AP7}

\begin{figure}[!t]
\centering \includegraphics[width=6.5in,totalheight=5in]{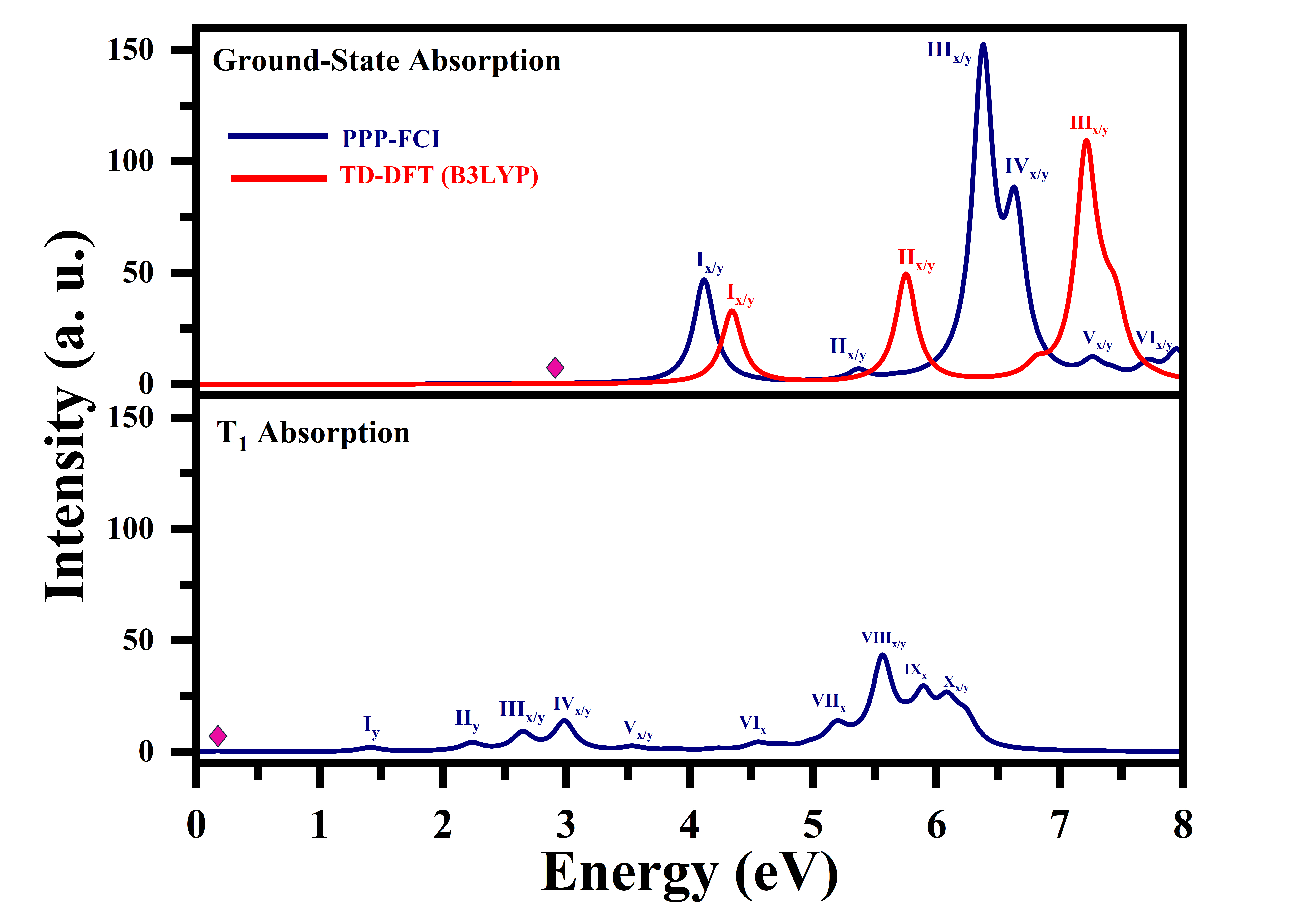}
\caption{The ground-state and triplet absorption spectra of AP7, i.e., Tri-s-triazine.
The rest of the information is same as that in the caption of Fig.
\ref{Fig.3.}.  }\label{Fig.11.}
\end{figure}

For AP7, we have observed a symmetry-induced doubly-degenerate $T_{1}$,
and $T_{1}\rightarrow T_{2}$ transition appear with $f=0.0026$ at
0.17 eV, with the $T_{2}$ wave function mainly consisting of the
$|H\rightarrow L\rangle$ single excitation, with subsidiary contribution
from double excitation $|H\rightarrow L;(H-1)_{1}\rightarrow(L+1)_{1}\rangle$,
$|H\rightarrow L;(H-1)_{2}\rightarrow(L+1)_{2}\rangle$, $|H\rightarrow(L+1)_{1};H\rightarrow(L+1)_{2}\rangle$,
and $|H\rightarrow L;H-2\rightarrow L\rangle$. Peak \RNum{1} (Fig.
~\ref{Fig.11.}) appears at 1.41 eV (Table S34) and the configuration
$|H-2\rightarrow L\rangle_{y}$ is a major contributor to the corresponding
excited state. The detailed analysis of the rest of the peaks is presented
in Table S34.

\ 

These results indicate that the onset of electronic absorption in
these molecules cannot be described solely by a single HOMO--LUMO
excitation. Instead, the low-energy absorption is dominated by correlated,
multi-configurational excited states arising from significant contributions
of several near-frontier occupied and virtual orbitals. Although the
dominant orbital configurations vary among the molecules, the first
intense absorption band\textcolor{blue}{{} }arises from mixtures of
several configurations involving near-frontier orbitals, highlighting
the importance of electron-correlation effects in shaping their optical
response.

\clearpage{}

\section{CONCLUSIONS}

Materials with \textcolor{black}{negative STGs} are interesting from
a fundamental point of view, as their first excited singlet state
lies below the first excited triplet state, violating the conventional
expectation based on electron exchange interaction as defined by Hund’s
rule of maximum multiplicity. This unusual energetic ordering may
facilitate singlet-dominated photophysical pathways and potentially
reduce triplet-related losses. To explore the $S_{1}-T_{1}$ inversion
in azaphenalenes, or poly-aza analogues of cyclazine, where nitrogen
atoms replace peripheral C-H groups, and to provide a detailed analysis
of their singlet and triplet spectra, we have investigated ten different
azaphenalene derivatives. This work employs the Pariser-Parr-Pople
(PPP) model, a semi-empirical, wave-function-based effective $\pi$-electron
Hamiltonian to consider the electron correlations in these conjugated
$\pi$-electron systems. Two schemes of long-range Coulomb interaction
(OHNO and MN) were considered within the PPP-model Hamiltonian. For
both the models of Coulomb interactions, we systematically tuned key
parameters including the on-site orbital energy, the Hubbard $U$,
and the hopping integrals to benchmark the model for reliably estimating
negative singlet-triplet gaps. In this work, we have employed six
sets of PPP parameters, i.e., three sets each of OHNO and MN models,
to explore the inverted STG and optical properties of the considered
molecules. Additionally, we carried out TDDFT calculations with various
exchange-correlation functionals (B3LYP, HSE06, CAM-B3LYP, PBE) to
test their efficacy in predicting the ST inversion, and found that
conventional adiabatic TDDFT methods failed to produce the negative
STG of these molecules. Then we performed FCI calculations on these
systems using the PPP model, and found that the method successfully
predicts the negative STG for OHNO-\RNum{2}, \RNum{3} and MN-\RNum{2},
\RNum{3} parameters. After comparing our results with the existing
experimental data of few azaphenalenes, we finally performed PPP-FCI
calculations using the OHNO-\RNum{3} parameters to compute the ground-state
and $T_{1}$ absorption spectra of these molecules to study their
optical response. Furthermore, we performed a detailed analysis of
the excited-state wave functions contributing to various peaks in
the absorption spectra, with the aim of understanding the influence
of electron-correlation effects on the optical properties of these
systems. We also compared our PPP-FCI ground-state absorption results
with those computed using the TDDFT method, and found that the TDDFT
peaks are significantly blue-shifted as compared to the PPP-FCI results.\textcolor{blue}{{}
}The present PPP-FCI results provide experimentally testable predictions
for the optical response of azaphenalenes, and may motivate future
spectroscopic investigations.

\section*{SUPPLEMENTARY MATERIAL}

The \textbf{supplementary material} contains additional structural, electronic, and optical data for the N-substituted phenalenyl molecules investigated in this work. It includes average ground-state bond lengths, HOMO--LUMO gaps obtained using different PPP Coulomb parameterizations and DFT, and vertical $S_{1}$ and $T_{1}$ excitation energies and singlet--triplet energy gaps calculated using different TDDFT functionals and PPP-FCI approaches. Detailed analyses of the ground-state and $T_{1}$ absorption spectra, including excitation energies, symmetries, transition dipole moments, oscillator strengths, polarization directions, and dominant many-body electronic configurations, are also provided for the investigated molecules.

\section*{ACKNOWLEDGMENTS}

M. R. gratefully acknowledges the Institute Postdoctoral Fellowship (IPDF) and the computational facilities provided by the Indian Institute of Technology Bombay.

\section*{AUTHOR DECLARATIONS}

\subsection*{Conflict of Interest}

The authors have no conflicts to disclose.

\section*{DATA AVAILABILITY}

The data that support the findings of this study are available
from the corresponding author upon reasonable request.


\bibliographystyle{apsrev4-2}
\bibliography{references}

\end{document}